\pdfoutput=1
\RequirePackage{ifpdf}
\ifpdf 
\documentclass[pdftex]{sigma}
\else
\documentclass{sigma}
\fi

\usepackage[all]{xy}

\def\isqrt#1{\sqrt{\smash{#1}\hbox{\vrule width0pt height 2ex depth 0.3ex}}}
\def\bigmathstrut{\hbox{\vrule width0pt height 2ex depth 1ex}}

\begin{document}

\newcommand{\arXivNumber}{1111.2027}

\allowdisplaybreaks

\renewcommand{\PaperNumber}{091}

\FirstPageHeading

\ShortArticleName{A Reciprocal Transformation for the Constant Astigmatism Equation}

\ArticleName{A Reciprocal Transformation\\
for the Constant Astigmatism Equation}

\Author{Adam HLAV\'A\v{C} and Michal MARVAN}

\AuthorNameForHeading{A.~Hlav\'a\v{c} and M.~Marvan}

\Address{Mathematical Institute in Opava, Silesian University in Opava,\\
Na Rybn\'\i\v{c}ku 1, 746 01 Opava, Czech Republic}
\Email{\href{mailto:Adam.Hlavac@math.slu.cz}{Adam.Hlavac@math.slu.cz}, \href{mailto:Michal.Marvan@math.slu.cz}{Michal.Marvan@math.slu.cz}}

\ArticleDates{Received May 07, 2014, in f\/inal form August 14, 2014; Published online August 25, 2014}

\Abstract{We introduce a~nonlocal transformation to generate exact solutions of the constant astigmatism equation
$z_{yy} + (1/z)_{xx} + 2 = 0$.
The transformation is related to the special case of the famous B\"acklund transformation of the sine-Gordon equation
with the B\"acklund parameter $\lambda = \pm1$.
It is also a~nonlocal symmetry.}


\Classification{53A05; 35A30; 35C05; 37K35}

\section{Introduction}
In this paper, we continue investigation of the \textit{constant astigmatism equation}
\begin{gather}
\label{CAE}
z_{yy} + \left(\frac{1}{z}\right)_{xx} + 2 = 0,
\end{gather}
which is the Gauss equation for \textit{constant astigmatism surfaces} immersed in the Euclidean space; see~\cite{B-M-I}.
These surfaces are def\/ined by the condition $\sigma - \rho = {\rm const}$, where~$\rho$,~$\sigma$ are the principal
radii of curvature and the constant is nonzero.
Without loss of generality, the ambient space may be scaled so that the constant is $\pm 1$, which is assumed in what
follows.
The same equation~\eqref{CAE} describes spherical \textit{orthogonal equiareal patterns}; see later in this section.

A brief history of constant astigmatism surfaces is included in~\cite{B-M-I, P-S} (apparently, they had no name
until~\cite{B-M-I}).
Ribaucour~\cite{Ri} and Bianchi~\cite{Bia-a,Bia-II} observed that evolutes (focal surfaces) of constant astigmatism
surfaces are pseudospherical (of constant negative Gaussian curvature).
Conversely, involutes of pseudospherical surfaces corresponding to parabolic geodesic nets are of constant astigmatism.
This yields a~pair of nonlocal transformations~\cite{B-M-I} between the constant astigmatism equation~\eqref{CAE} and
the integrable sine-Gordon equation
\begin{gather}
\label{sG}
u_{\alpha\beta} = \sin u,
\end{gather}
which is the Gauss equation for pseudospherical surfaces in terms of asymptotic Chebyshev coordinates~$\alpha$,~$\beta$.
Hence, the constant astigmatism equation is also integrable; see~\cite{B-M-I} for its zero curvature representation.
Moreover, the famous Bianchi superposition principle for the sine-Gordon equation can be extended in such a~way that an
arbitrary number of solutions of the constant astigmatism equation can be obtained by purely algebraic manipulations and
dif\/ferentiation; see~\cite{H-M}.

The class of sine-Gordon solutions that can serve as a~seed is limited, since the initial step involves integration of
the nonlinear (although linearisable) equations~\eqref{BT} below.
The class contains all multisoliton solutions, which themselves can be generated from the zero seed.
Another successfully planted seed we know of is the travelling wave used by Hoenselaers and Miccich\`e~\cite{Ho-Mi}.

In this paper, we look for another solution-generating tool that would not require solving dif\/ferential equations.
We introduce two (interrelated) auto-transformations $\mathcal X_A$ and $\mathcal Y_B$ that, in geometric terms,
correspond to taking the involute of the evolute.
Each generates a~three-parametric family of solutions from a~single seed, but when applied in combination, they have an
unlimited generating power in terms of the number of arbitrary parameters in the solution.

The transformations $\mathcal X_A$ and $\mathcal Y_B$ are B\"acklund transformations {\it sensu}
B\"acklund~\cite{Bae,Gour}, since each is determined by four relations of no more than the f\/irst order (although modern
usage often sees this term as implying that independent variables are preserved, the original meaning is as stated).
We call $\mathcal X_A$ and $\mathcal Y_B$ {\it reciprocal transformations} since, up to point transformations, $\mathcal
X_A$ and $\mathcal Y_B$ are equivalent to~$\mathcal X$ and~$\mathcal Y$ satisfying
\begin{gather*}
\mathcal X^2 = \mathcal Y^2 = {\rm Id},
\end{gather*}
which is a~property characteristic of reciprocal transformations~\cite{K-R}.
For the history and overview of reciprocal transformations and their wide applications in physics and geometry
see~\cite[Chapter~3]{R-Sh} and \cite[Section~6.4]{R-S}.
Reciprocal invariants, linked to invariants in Lie sphere geometry, are available in the context of hydrodynamic-type
systems~\cite{Fer-I,Fer-II,F-R-S}.
Geometry of immersed surfaces is rich in nonlocal transformations to the sine-Gordon equation, see,
e.g.,~\cite[\S~3.3]{R-S-S}; this example is of particular interest in the context of our present ef\/forts
(compare~\cite[equation~(3.27)]{R-S-S} to~\cite[Table~1, row~6b]{B-M-II}).
Nonlocal transformations between general integrable equations are often reciprocal or decomposable into a~chain where
one of the factors is reciprocal; this extends to hierarchies, see~\cite{R-W} and~\cite[\S~6.4]{R-S-S}.

Transformations $\mathcal X_A$ and $\mathcal Y_B$ only depend on the computation of path-independent line integrals,
which puts lower demands on the seeds.
The sine-Gordon equation is bypassed and the transformations are immediately applicable to solutions of the constant
astigmatism equation with no apriori given sine-Gordon counterpart, such as the Lipschitz solution~\cite{H-M-2, Lip}.
If the seeds are given in parametric form, then so are the generated solutions.

Our work would be incomplete without explicitly constructing the transformed surface of constant astigmatism.
To obtain compact formulas, a~small but useful digression is made.
According to~\cite{H-M}, to every surface of constant astigmatism there corresponds an orthogonal equiareal pattern on
the Gaussian sphere; the same conclusion was made by Bianchi~\cite[\S~375, equation~(20)]{Bia-II} in the context of
pseudospherical congruences.
By an orthogonal equiareal pattern~\cite{H-M,Sad2, Sad1} we mean a~parameterization such that the metric assumes the form
\begin{gather*}
z\,{\mathrm d} x^2 + \frac1z\, {\mathrm d} y^2
\end{gather*}
(which is, incidentally, of relevance to two-dimensional plasticity under the Tresca yield condition,~\cite{Sad2}).
The name ref\/lects the fact that ``uniformly spaced'' coordinate lines $x = a_1 t + a_0$ and $y = b_1 s + b_0$, where
$a_1,a_0,b_1,b_0 = {\rm const}$, form equiareal curvilinear rectangles.

The contents of this paper are as follows.
In Section~\ref{sect:pointsymm} we recall symmetries and conservation laws of the constant astigmatism equation.
Section~\ref{sect:Constr} contains a~derivation of the reciprocal transformations.
Starting with a~constant astigmatism surface we construct its pseudospherical evolute and then a~family of new parallel
surfaces of constant astigmatism.
New solutions of the constant astigmatism equation then result from f\/inding curvature coordinates on these surfaces.
Section~\ref{section4} summarizes the transformations obtained in the preceding section.
In Section~\ref{sect:nloc} we obtain the same transformations as nonlocal symmetries of the constant astigmatism
equation.
In Section~\ref{sect:sG} we show that they correspond to the B\"acklund transformation for the sine-Gordon equation with
the B\"acklund parameter $\lambda = \pm1$.
In Section~\ref{sect:OEP} we describe the transformations in terms of the constant astigmatism surfaces and the
orthogonal equiareal patterns on the Gaussian sphere.
Finally, the last section contains several exact solutions.

\section{Point symmetries}
\label{sect:pointsymm}

According to~\cite{B-M-I}, there are three independent continuous Lie symmetries of equation~\eqref{CAE}:
the~$x$-translation
\begin{gather*}
{\mathcal T}^x_a (x, y, z) = (x + a, y, z),
\qquad
a \in {\mathbb R},
\end{gather*}
the~$y$-translation
\begin{gather*}
{\mathcal T}^y_b (x, y, z) = (x, y + b, z),
\qquad
b \in {\mathbb R},
\end{gather*}
and the scaling
\begin{gather*}
{\mathcal S}_c (x, y, z) = \big(x/c, c y, c^2 z\big),
\qquad
c \in {\mathbb R} \setminus \{0\}.
\end{gather*}
The known discrete symmetries are exhausted by the involution (or duality)
\begin{gather*}
{\mathcal J}(x, y, z) = \left(y, x, \frac 1 z\right),
\end{gather*}
the~$x$-reversal ${\mathcal R}^x(x, y, z) = (-x, y, z)$, and the~$y$-reversal ${\mathcal R}^y(x, y, z) = (x, -y, z)$.
To avoid possible misunderstanding, we stress that ${\mathcal T}^x$, ${\mathcal T}^y$, ${\mathcal R}^x$, ${\mathcal
R}^y$ should be understood as single symbols, similarly to ${\mathcal S}$, ${\mathcal J}$.
The superscripts~$x$,~$y$ refer to the af\/fected position in the triple $(x,y,z)$.

Obviously,
\begin{alignat*}{5}
& {\mathcal J} \circ {\mathcal J} = {\rm Id},
\qquad &&
{\mathcal J} \circ {\mathcal T}^x_a = {\mathcal T}^y_{a} \circ {\mathcal J},
\qquad &&
{\mathcal J} \circ {\mathcal T}^y_a = {\mathcal T}^x_{a} \circ {\mathcal J}, && &
\\
& {\mathcal S}_c \circ {\mathcal T}^x_a = {\mathcal T}^x_{a/c} \circ {\mathcal S}_c,
\qquad &&
{\mathcal S}_c \circ {\mathcal T}^y_b = {\mathcal T}^y_{cb} \circ {\mathcal S}_c,
\qquad &&
{\mathcal S}_c \circ {\mathcal J} = {\mathcal J} \circ {\mathcal S}_{1/c},
\qquad &&
{\mathcal R}^x \circ {\mathcal S}_{-1} = {\mathcal R}^y.&
\end{alignat*}
Translations and reversals correspond to mere reparameterizations of the constant astigmatism surfaces.
The scaling symmetry takes a~surface to a~{\it parallel surface}, obtained when moving every point of the surface
a~constant distance along the normal ({\it offsetting}).
The involution swaps the orientation, interchanges~$x$ and~$y$, and makes a~unit of\/fsetting.
Solutions invariant with respect to the local Lie symmetries can be found in~\cite[Proposition~1]{H-M-2}; they
correspond to the Lipschitz~\cite{Lip} class of constant astigmatism surfaces.
Higher order symmetries have been considered in~\cite{B-M-I} and~\cite{P-Z}; they will not be needed in this paper.

We shall also need the six f\/irst-order conservation laws of equation~\eqref{CAE}, which are easy to compute following,
e.g.,~\cite{B-V-V}.
The associated six potentials~$\chi$,~$\eta$, $\xi$,~$\theta$,~$\alpha$,~$\beta$ satisfy
\begin{alignat}
{3} & \chi_x = z_y + y,
\qquad
&& \chi_y = \frac{z_x}{z^2} - x, &
\nonumber
\\
&\eta_x = x z_y,
\qquad
&& \eta_y = x \frac{z_x}{z^2} + \frac1z - x^2, &
\nonumber
\\
& \xi_x = -y z_y + z - y^2,
\qquad
&& \xi_y = -y \frac{z_x}{z^2}, &
\nonumber
\\
& \theta_x = x y z_y - x z + \frac12 x y^2,
\qquad
&& \theta_y = x y \frac{z_x}{z^2} + \frac y z - \frac12 x^2 y &
\label{potentials}
\end{alignat}
and
\begin{alignat}
{3} & \alpha_x = \frac{\sqrt{(z_x + z z_y)^2 + 4 z^3}}{4z},
\qquad
&& \alpha_y = \frac{\sqrt{(z_x + z z_y)^2 + 4 z^3}}{4z^2}, &
\nonumber
\\
& \beta_x = \frac{\sqrt{(z_x - z z_y)^2 + 4 z^3}}{4z},
\qquad
&& \beta_y = -\frac{\sqrt{(z_x - z z_y)^2 + 4 z^3}}{4z^2}.&
\label{alphabeta}
\end{alignat}
Equations~\eqref{potentials}, \eqref{alphabeta} are compatible by virtue of equation~\eqref{CAE}.
It is not a~pure coincidence that the same symbols $\alpha$, $\beta$ occur in equations~\eqref{sG} and~\eqref{alphabeta},
see Section~\ref{sect:sG} below.
Assuming~$z$ positive in accordance to its geometrical meaning~\cite{B-M-I,H-M}, the radicands in~\eqref{alphabeta} are
positive as well.
On the other hand, Manganaro and Pavlov~\cite{M-P} considered the class of solutions such that one of the two radicands
is zero.

The involution ${\mathcal J}$ acts on the potentials as follows: $\eta \leftrightarrow \xi$, while $\chi \to -\chi$,
$\theta \to -\theta$, $\alpha \leftrightarrow \alpha$, and $\beta \leftrightarrow -\beta$.
The action of the other symmetries is considered below.

\section{A geometric construction}
\label{sect:Constr}

Let $z(x,y)$ be a~solution of the constant astigmatism equation~\eqref{CAE}.
Under the choice of scale such that $\sigma - \rho = 1$, the fundamental forms of the corresponding surface of constant
astigmatism are
\begin{gather*}
\mathbf{I} = u^2\, {\mathrm d} x^2 + v^2 \,{\mathrm d} y^2,
\qquad
\mathbf{II} = \frac{u^2}\rho\, {\mathrm d} x^2 + \frac{v^2}\sigma\, {\mathrm d} y^2,
\end{gather*}
where
\begin{gather*}
u = \frac{\ln z-2}{2} \sqrt z,
\qquad
v = \frac{\ln z}{2 \sqrt z},
\qquad
\rho = \frac{\ln z-2}{2},
\qquad
\sigma = \frac{\ln z}{2},
\qquad
\end{gather*}
see~\cite{B-M-II}.
Obviously,
\begin{gather}
\label{adapted}
u v = \rho \sigma.
\end{gather}
Once $\sigma - \rho = \pm1$, equation~\eqref{adapted} means that~$x$,~$y$ are the {\it adapted} curvature coordinates in
the sense of~\cite[Definition~1]{H-M} (they are also {\it normal} coordinates in the sense of~\cite{G-M}).

The corresponding surface $\mathbf{r}(x,y)$ of constant astigmatism and its unit normal $\mathbf n(x,y)$ satisfy the
Gauss--Weingarten system
\begin{gather}
\mathbf{r}_{xx} = \frac{(\ln z) z_x}{2 (\ln z - 2) z} \mathbf{r}_x - \frac{(\ln z - 2) z z_y}{2 \ln z} \mathbf{r}_y +
\frac12 (\ln z - 2) z \mathbf{n},
\nonumber
\\
\mathbf{r}_{xy} = \frac{(\ln z) z_y}{2 (\ln z - 2) z} \mathbf{r}_x - \frac{(\ln z - 2) z_x}{2 z \ln z} \mathbf{r}_y,
\nonumber
\\
\mathbf{r}_{yy} = \frac{(\ln z) z_x}{2 (\ln z - 2) z^3} \mathbf{r}_x - \frac{(\ln z - 2) z_y}{2 z \ln z} \mathbf{r}_y +
\frac{\ln z}{2 z} \mathbf{n},
\nonumber
\\
\mathbf{n}_x = -\frac{2}{\ln z - 2} \mathbf{r}_x,
\qquad
\mathbf{n}_y = -\frac{2}{\ln z} \mathbf{r}_y,
\label{GW}
\end{gather}
which is compatible as a~consequence of equation~\eqref{CAE}.

Let us construct the pseudospherical evolute~\cite{Bia-a} of the surface $\mathbf{r}$.
By def\/inition, the evolute has two sheets formed by the loci of the principal centres of curvature.
We choose one of the two evolutes, given~by
\begin{gather*}
\hat{\mathbf{r}} = \mathbf{r} + \sigma \mathbf{n},
\qquad
\hat{\mathbf{n}} = \frac{\mathbf{r}_y}{v} = \frac{2 \sqrt z}{\ln z} \mathbf{r}_y.
\end{gather*}
The f\/irst fundamental form of this evolute is
\begin{gather*}
\hat{\mathbf{I}} = \frac {4{z}^{3}+ z_x^{2}}{4{z}^{2}} \, {\mathrm d} x^2 + \frac {z_x z_y}{2{z}^{2}} \, {\mathrm d} x
\,{\mathrm d} y + \frac {z_y^{2}}{4{z}^{2}} \, {\mathrm d} y^2 = z \, {\mathrm d} x^2 + \frac{{\mathrm d} z^2}{4 z^2}.
\end{gather*}
One easily sees that the Gauss curvature is $-1$, as expected.

Next we construct the involute to this pseudospherical surface in order to obtain a~new surface of constant astigmatism
together with a~new solution of the equation~\eqref{CAE}.
Following~\cite[\S~136]{Bia-I} (see also~\cite{Wea}), we let~$X$ and~$Y$ be parabolic geodesic coordinates on the
pseudospherical surface.
By def\/inition, the f\/irst fundamental form should be
\begin{gather*}
\hat{\mathbf{I}} = {\mathrm d} X^2 + {\mathrm e}^{2X} {\mathrm d} Y^2.
\end{gather*}
Comparing the coef\/f\/icients, we obtain
\begin{gather*}
X_x^2 + {\mathrm e}^{2X}Y_x^2 = z + \frac{z_x^2}{4z^2},
\qquad
2 X_x X_y + 2 {\mathrm e}^{2X} Y_x Y_y = \frac{z_x z_y}{2z^2},
\qquad
X_y^2 + {\mathrm e}^{2X} Y_y^2 = \frac{z_y^2}{4z^2}.
\end{gather*}
Solving the last two equations for $Y_x$, $Y_y$, we have
\begin{gather}
\label{cov23}
Y_x = \frac{z_x z_y - 4 z^2 X_x X_y} {2 z {\mathrm e}^X \isqrt{z_y^2 - 4 z^2 X_y^2}},
\qquad
Y_y = \frac{\isqrt{z_y^2 - 4 z^2 X_y^2}\bigmathstrut}{2 z {\mathrm e}^X},
\end{gather}
which allows us to convert the remaining equation into
\begin{gather}
\label{cov1}
X_x = \frac{z_x X_y + \isqrt{z z_y^2 - 4 z^3 X_y^2}\bigmathstrut}{z_y}.
\end{gather}
Substituting into~\eqref{cov23} and performing cross-dif\/ferentiation, we obtain
\begin{gather}
\label{Ric}
X_{yy} = -X_y^2 + \frac{z z_{xx} - 2 z_x^2 - z^2 z_y^2 - 2 z^3}{z^3 z_y} X_y + \frac{z_y^2}{4 z^2}.
\end{gather}
Now the system consisting of equations~\eqref{cov23},~\eqref{cov1} and~\eqref{Ric} is compatible by virtue of
equation~\eqref{CAE}.

The involute we look for is given~by
\begin{gather}
\tilde{\mathbf{r}} = \hat{\mathbf{r}} + (a - X) \hat{\mathbf{r}}_X = \mathbf{r} + \left(\frac{1}{2} \ln z + {\frac
{2z(a-X) X_y}{z_y}}\right)\mathbf{n} + 2 (a-X) \frac{\isqrt {z_y^{2} - 4 {z}^{2} X_y^{2}}\bigmathstrut}{\sqrt{z} z_y (2 - \ln z)}
\mathbf{r}_x,
\label{invo}
\end{gather}
where~$a$ is an arbitrary constant.
The unit normal vector to the involute is
\begin{gather*}
\tilde{\mathbf{n}} = \hat{\mathbf{r}}_X = \frac{2 z X_y}{z_y}\mathbf{n} - \frac{2\isqrt {z_y^{2} - 4 {z}^{2}
X_y^{2}}\bigmathstrut}{\sqrt{z} z_y (\ln z - 2)} \mathbf{r}_x.
\end{gather*}
To obtain~$X$, we have to solve the compatible system~\eqref{cov1},~\eqref{Ric}.
The other unknown~$Y$ is no more needed.

The system~\eqref{cov1},~\eqref{Ric} has the obvious particular solution
\begin{gather*}
X_0 = \tfrac{1}{2}\ln z + c_2,
\end{gather*}
which corresponds to the constant astigmatism surface $\tilde{\mathbf{r}} = \mathbf{r} + (a - c_2) \mathbf{n}$.
Thus, we recover the constant astigmatism surface we started with along with all its parallel surfaces.

To f\/ind the general solution of the system~\eqref{cov1},~\eqref{Ric}, we f\/irst observe that~\eqref{Ric} is a~Riccati
equation in~$X_y$.
Knowing one particular solution $X_{0,y}$ is suf\/f\/icient for f\/inding the general solution $X_y$, see, e.g.,~\cite{Po-Za}.
Omitting the details, we present the general solution
\begin{gather}
\label{gensol}
X = \ln \frac{(x+c_1)^2 z + 1}{\sqrt z} + c_2
\end{gather}
of the system~\eqref{cov1},~\eqref{Ric}.
In order to simplify the formulas below, we remove the integration constant $c_1$ by reparameterization ${\mathcal
T}^x_{c_1}$.
Then
\begin{gather*}
X = \ln \frac{x^2 z + 1}{\sqrt{z}} + c_2,
\end{gather*}
which, if substituted into formula~\eqref{invo}, yields the family of involutes
\begin{gather}
\label{invol}
\tilde{\mathbf{r}} = \mathbf{r} + \left(\frac{x^2 z \ln z}{x^2 z + 1} + \frac{x^2 z - 1}{x^2 z + 1} \big(\ln\big(x^2 z +
1\big) + a\big)\right) \mathbf{n} + 2x\frac{2a - 2\ln\big(x^2 z + 1\big) + \ln z}{\big(x^2 z + 1\big)(2 - \ln z)}
\mathbf{r}_x,
\end{gather}
where $c_2$ has been absorbed into~$a$.
The corresponding unit normal is
\begin{gather}
\label{normal}
\tilde{\mathbf{n}} = \frac{x^2 z - 1}{x^2 z + 1}\mathbf{n} + \frac{4x}{\big(x^2 z + 1\big)(2 - \ln z)} \mathbf{r}_x.
\end{gather}
A~routine computation (see below) shows that the surface $\tilde{\mathbf{r}}(x,y)$ has a~constant astigmatism, and so
has $\tilde{\mathbf{r}} (x+c_1,y)$, which corresponds to~$X$ given by the general solution~\eqref{gensol}.
The para\-me\-ter~$a$ corresponds to the of\/fsetting, meaning a~parallel surface.

However, one more step is required in order to f\/ind the corresponding solution of equation~\eqref{CAE}.
Namely, we have to f\/ind the adapted curvature coordinates~$x'$, $y'$ for the involute.
In order that~$x'$,~$y'$ be curvature coordinates, $\partial/\partial x'$ and $\partial/\partial y'$ have to be
eigenvectors of the shape operator.
The shape operator is too lengthy to be written here, but its eigenvalues (principal curvatures)~$1/\rho'$,~$1/\sigma'$
are simple enough (cf.\ equation~\eqref{sigmaprime} below).
Computing the eigenvectors, and choosing an assignment between $\partial/\partial x'$, $\partial/\partial y'$ and the
two eigenvectors, we obtain
\begin{gather}
\label{xyprime curv}
x'_y = \frac{x z_y}{x z_x - x^2 z^2 + z} x'_x,
\qquad
y'_y = \frac{x z_x - x^2 z^2 + z}{x z^2 z_y} y'_x.
\end{gather}
Under conditions~\eqref{xyprime curv}, we have
\begin{gather*}
\tilde{\mathbf{I}} = u^{\prime2} \, {\mathrm d} x^{\prime2} + v^{\prime2}\, {\mathrm d} y^{\prime2},
\qquad
\tilde{\mathbf{II}} = \frac{u^{\prime2}}{\rho'} \, {\mathrm d} x^{\prime2} + \frac{v^{\prime2}}{\sigma'} \, {\mathrm d}
y^{\prime2},
\end{gather*}
where
\begin{alignat}
{3} & \rho' = \ln\frac{\sqrt z}{x^2 z + 1} + a + 1,
\qquad
&& \sigma' = \ln\frac{\sqrt z}{x^2 z + 1} + a,&
\nonumber
\\
& u' = \frac{x z_x - x^2 z^2 + z}{(x^2 z + 1) \sqrt z} \frac{\rho'}{x'_x},
\qquad
&& v' = \frac{x \sqrt z z_y}{x^2 z + 1} \frac{\sigma'}{y'_x}.&
\label{sigmaprime}
\end{alignat}
We immediately see that $\rho' - \sigma' = 1$, which implies that the surface $\tilde{\mathbf{r}}$ is of constant
astigmatism as required.

In order that $x'$, $y'$ be adapted in the sense of~\cite[Definition~1]{H-M}, we require $u' v' = \pm \rho' \sigma'$,
which implies
\begin{gather}
\label{xyprime adapt yx}
y'_x = \pm \frac{x z_x - x^2 z^2 + z}{\big(x^2 z + 1\big)^2} \frac{x z_y}{x'_x}.
\end{gather}
The second equation~\eqref{xyprime curv} becomes
\begin{gather}
\label{xyprime adapt yy}
y'_y = \pm \left(\frac{x z_x - x^2 z^2 + z}{\big(x^2 z + 1\big) z}\right)^2 \frac1{x'_x}.
\end{gather}
By cross-dif\/ferentiation between equations~\eqref{xyprime adapt yx} and~\eqref{xyprime adapt yy}, we obtain the
second-order equation
\begin{gather*}
x'_{xx} = \frac{x \big(x^2 z + 1\big) z_{xx} - 2 x^3 z_x^2 - 6 x^2 z z_x + 2 z_x + 2 x^3 z^3 - 6 x z^2} {\big(x^2 z +
1\big)\big(x z_x - x^2 z^2 + z\big)} x'_x,
\end{gather*}
possessing the general solution
\begin{gather*}
x'_x = b \frac{x z_x - x^2 z^2 + z}{\big(x^2 z + 1\big)^2},
\end{gather*}
where~$b$ denotes an arbitrary function of~$y$.
Inserting into~\eqref{xyprime curv},~\eqref{xyprime adapt yx} and~\eqref{xyprime adapt yy}, we obtain
\begin{gather*}
x'_y = b \frac{x z_y}{(x^2 z + 1)^2},
\qquad
y'_x = \pm \frac1b x z_y,
\qquad
y'_y = \pm \frac1b \frac{x z_x - x^2 z^2 + z}{z^2}.
\end{gather*}
By cross-dif\/ferentiation between the two equations on~$y'$, we receive $b_y = 0$, meaning that~$b$ is a~constant.
The last four equations on~$x'$ and~$y'$ are compatible now; their general solution is simply
\begin{gather*}
x' = b \frac{x z}{x^2 z + 1} + c_2,
\qquad
y' = \pm \frac 1b \eta + c_3,
\end{gather*}
where
\begin{gather*}
{\mathrm d}\eta = x z_y \, {\mathrm d} x + \left(x \frac{z_x}{z^2} + \frac1z - x^2\right) {\mathrm d} y
\end{gather*}
as introduced in Section~\ref{sect:pointsymm} above.
Finally,
\begin{gather*}
\tilde{\mathbf{III}} = \frac1{b^2} \frac{\big(x^2 z + 1\big)^2} z \, {\mathrm d} x^{\prime2} + b^2 \frac{z}{\big(x^2 z +
1\big)^2}\,  {\mathrm d} y^{\prime2}.
\end{gather*}
Hence,
\begin{gather*}
z' = \frac1{b^2} \frac{\big(x^2 z + 1\big)^2} z.
\end{gather*}
Recall that we applied the translation ${\mathcal T}^x_{c_1}$ after equation~\eqref{gensol} in order to remove the
parameter~$c_1$.
Reintroducing~$c_1$, which amounts to replacing~$x$ with $x + c_1$ in the above formulas, we obtain
\begin{gather}
x' = b \frac{(x + c_1) z}{(x + c_1)^2 z + 1} + c_2,
\qquad
y' = \pm \frac 1b \big(\eta + c_1 \chi - c_1 x y - c_1^2 y\big) + c_3,
\nonumber
\\
z' = \frac1{b^2} \frac{((x + c_1)^2 z + 1)^2} z,
\label{XA}
\end{gather}
where~$\chi$ was introduced in Section~\ref{sect:pointsymm} above.
The parameters $c_i$ can be conveniently encoded in a~matrix, see Proposition~\ref{prop:A} below.

To sum up, we started with a~constant astigmatism surface~$\mathbf r$, constructed its pseudospherical
image~$\hat{\mathbf r}$, then reconstructed the full preimage $\tilde{\mathbf r}$, ref\/lecting the freedom of choice of
the parabolic geodesic system on~$\hat{\mathbf r}$.
Accordingly, it should not come as a~surprise that the transformation has a~limited generating power, measured by the
number of arbitrary constants the resulting solution depends on.
The generating power becomes unlimited only if two such transformations, using dif\/ferent sheets of the evolute, are
combined.
This will be discussed in Section~\ref{sect:sG} below.

\section{The reciprocal transformations and their properties}
\label{section4}

Consider the formulas~\eqref{XA}.
Setting all integration constants $c_i$ to zero,~$b$ to~$1$, and choosing the `$+$' sign, we obtain a~transformation
$\mathcal X(x,y,z) = (x',y',z')$, def\/ined~by
\begin{gather}
\label{RTX}
x' = \frac{x z}{x^2 z + 1},
\qquad
y' = \eta,
\qquad
z' = \frac{\big(x^2 z + 1\big)^2} z.
\end{gather}
Using conjugation with the involution ${\mathcal J}$, we obtain another transformation $\mathcal Y(x,y,z) = (x^*$,
$y^*,z^*)$, where
\begin{gather}
\label{RTY}
x^* = \xi,
\qquad
y^* = \frac{y}{y^2 + z},
\qquad
z^* = \frac{z}{\big(y^2 + z\big)^2}.
\end{gather}
Formulas~\eqref{RTY} follow from formulas~\eqref{RTX} and the relation ${\mathcal Y} = {\mathcal J} \circ {\mathcal X}
\circ {\mathcal J}$.

\begin{remark}
\label{rem:nonuniq}
We remind the reader that $\xi$ and~$\eta$ are potentials def\/ined in Section~\ref{sect:pointsymm}.
Therefore, they are unique up to an integration constant, which is not to be neglected, because it represents
a~parameter in the solution.
Alternatively speaking, symbols ${\mathcal X}$ and ${\mathcal Y}$ can be viewed as standing for the compositions
${\mathcal T}^y_b \circ {\mathcal X}$ and ${\mathcal T}^x_a \circ {\mathcal Y}$, respectively, where $a,b$ are arbitrary
constants.
\end{remark}

\begin{proposition}
Let $z(x,y)$ be a~solution of the constant astigmatism equation~\eqref{CAE}, $\xi$,~$\eta$ the corresponding
potentials~\eqref{potentials}.
Let ${\mathcal X}(x,y,z) = (x',y',z')$ and ${\mathcal Y}(x,y,z) = (x^*,y^*,z^*)$ be determined by formulas~\eqref{RTX}
and~\eqref{RTY}.
Then $z'(x',y')$ and $z^*(x^*,y^*)$ are solutions of the constant astigmatism equation~\eqref{CAE} as well.
\end{proposition}

\begin{proof}
The statement follows from the reasoning in the preceding section.
A~routine, straightforward, but cumbersome proof consists in computing $z'_{x'x'} + (1/z')_{y'y'} + 2 = 0$.
\end{proof}

Let us note that the f\/irst derivatives transform according to the formulas
\begin{gather*}
\left(
\begin{matrix}
z'_{x'}
\\
z'_{y'}
\end{matrix}
\right) = J_{{\mathcal X}}^{-1} \left(
\begin{matrix}
z'_{x}
\\
z'_{y}
\end{matrix}
\right),
\qquad
\left(
\begin{matrix}
z^*_{x^*}
\\
z^*_{y^*}
\end{matrix}
\right) = J_{{\mathcal Y}}^{-1} \left(
\begin{matrix}
z^*_{x}
\\
z^*_{y}
\end{matrix}
\right),
\end{gather*}
where $J_{{\mathcal X}}$ and $J_{{\mathcal Y}}$ are the Jacobi matrices
\begin{gather*}
J_{{\mathcal X}} = \left(
\begin{matrix}
x'_x & y'_x
\\
x'_y & y'_y
\end{matrix}
\right) = \left(
\begin{matrix}
\dfrac{x z_x - x^2 z^2 + z}{\big(x^2 z+1\big)^2} & x z_y \vspace{1mm}
\\
\dfrac{x z_y}{\big(x^2 z+1\big)^2} & \dfrac{x z_x - x^2 z^2 + z}{z^2}
\end{matrix}
\right),
\\
J_{{\mathcal Y}} = \left(
\begin{matrix}
x^*_x & y^*_x
\\
x^*_y & y^*_y
\end{matrix}
\right) = \left(
\begin{matrix}
-y z_y + z - y^2 & -\dfrac{y z_x}{\big(z + y^2\big)^2} \vspace{1mm}
\\
-\dfrac{y z_x}{z^2} & -\dfrac{y z_y - z + y^2}{\big(z + y^2\big)^2}
\end{matrix}
\right).
\end{gather*}
Formulas for the second derivatives are too lengthy to be printed.

\begin{proposition}
Under a~suitable choice of integration constants, ${\mathcal X} \circ {\mathcal X} = {\rm Id}$ and ${\mathcal Y} \circ
{\mathcal Y} = {\rm Id}$.
\end{proposition}

\begin{proof}
It is straightforward to see that $x'' = x' z'/(x^{\prime2} z' + 1) = x$ and $z'' = (x^{\prime2} z' + 1)^2/z' = z$.
Let us compute $y'' = \eta'$, omitting technical details.
According to~\eqref{potentials}, $\eta'$ is def\/ined~by
\begin{gather*}
\eta'_{x'} = x' z'_{y'} = -\frac{(x^2 z + 1)^2 x z^2 z_y} {\big(x z_x - x^2 z^2 + z + x z z_y\big)\big(x z_x - x^2 z^2 +
z - x z z_y\big)},
\\
\eta'_{y'} = x' \frac{z'_{x'}}{z^{\prime2}} + \frac1{z'} - x^{\prime2} = \frac{\big(x z_x - x^2 z^2 + z\big) z^2}
{\big(x z_x - x^2 z^2 + z + x z z_y\big)\big(x z_x - x^2 z^2 + z - x z z_y\big)}.
\end{gather*}
Therefore,
\begin{gather*}
\eta'_x = \eta'_{x'} x'_x + \eta'_{y'} y'_x = \eta'_{x'} \frac{x z_x - x^2 z^2 + z}{\big(x^2 z + 1\big)^2} + \eta'_{y'}
x z_y = 0,
\\
\eta'_y = \eta'_{x'} x'_y + \eta'_{y'} y'_y = \eta'_{x'} \frac{x z_y}{\big(x^2 z + 1\big)^2} + \eta'_{y'} \frac{x z_x -
x^2 z^2 + z}{z^2} = 1.
\end{gather*}
Suppressing the integration constants, we obtain $y'' = \eta' = y$.
\end{proof}

Because of this property, ${\mathcal X}$ and ${\mathcal Y}$ are called {\it reciprocal transformations}, see,
e.g.,~\cite{K-R} and~\cite{R-S}.

\begin{remark}
\label{rem:circinversion}
The transformation ${\mathcal X}$ admits a~restriction to the variables~$x$,~$z$ and then
\begin{gather*}
x^{\prime2} z' = x^2 z,
\qquad
\big(x^{\prime2} + 1/z'\big) \big(x^2 + 1/z\big) = 1.
\end{gather*}
Therefore, ${\mathcal X}$ can be identif\/ied with the circle inversion in the $(x, z^{-1/2})$-subspace.
Similarly, ${\mathcal Y}$~admits a~restriction to the variables~$y$,~$z$, and then
\begin{gather*}
y^{\prime2}/z' = y^2/z,
\qquad
\big(y^{\prime2} + z'\big) \big(y^2 + z\big) = 1.
\end{gather*}
In this case, we obtain the circle inversion in the $(y, z^{1/2})$-subspace.
\end{remark}

The following identities are obvious:
\begin{gather*}
{\mathcal X} \circ {\mathcal J} = {\mathcal J} \circ {\mathcal Y},
\qquad
{\mathcal X} \circ {\mathcal S}_c = {\mathcal S}_{1/c} \circ {\mathcal X},
\qquad
{\mathcal Y} \circ {\mathcal S}_c = {\mathcal S}_{1/c} \circ {\mathcal Y}.
\end{gather*}
Slightly abusing the notation, we have also
\begin{gather*}
{\mathcal X} \circ {\mathcal T}^y_b = {\mathcal X} = {\mathcal T}^y_b \circ {\mathcal X},
\qquad
{\mathcal Y} \circ {\mathcal T}^x_a = {\mathcal Y} = {\mathcal T}^x_a \circ {\mathcal Y}.
\end{gather*}
There is no similar identity for ${\mathcal X} \circ {\mathcal T}^x_a$ and ${\mathcal Y} \circ {\mathcal T}^y_b$.
Instead, ${\mathcal X}$, ${\mathcal T}^x_a$ generate a~three-parameter group, and so do ${\mathcal Y}$, ${\mathcal
T}^y_b$.

\begin{proposition}
\label{prop:A}
Let $z(x,y)$ be a~solution of the constant astigmatism equation~\eqref{CAE}, $\chi,\eta,\xi$ the corresponding
potentials~\eqref{potentials}, and
\begin{gather*}
A = \left(
\begin{matrix}
a_{11} & a_{12}
\\
a_{21} & a_{22}
\end{matrix}
\right)
\end{gather*}
a~real matrix such that $\det A = \pm 1$.
Let ${\mathcal X}_A(x,y,z) = (x_A',y_A',z_A')$ and ${\mathcal Y}_A(x,y,z) = (x_A^*,y_A^*,z_A^*)$, where
\begin{gather*}
x_A' = \frac{(a_{11} + a_{12} x) (a_{21} + a_{22} x) z + a_{12} a_{22}} {(a_{11} + a_{12} x)^2 z + a_{12}^2},
\qquad
\pm y_A' = a_{12}^2 \eta + a_{11} a_{12} \chi - a_{11} y(a_{11} + a_{12} x),
\\
z_A' = \frac{\big((a_{11} + a_{12} x)^2 z + a_{12}^2\big)^2} z
\end{gather*}
and
\begin{gather*}
\pm x_A^* = a_{12}^2 \xi - a_{11} a_{12} \chi - a_{11} x (a_{11} + a_{12} y),
\qquad
y_A^* = \frac{(a_{11} + a_{12} y) (a_{21} + a_{22} y) + a_{12} a_{22} z} {(a_{11} + a_{12} y)^2 + a_{12}^2 z},
\\
z_A^* = \frac z {\big((a_{11} + a_{12} y)^2 + a_{12}^2 z\big)^2}.
\end{gather*}
Then $z_A'(x_A',y_A')$ and $z_A^*(x_A^*,y_A^*)$ are solutions of the constant astigmatism equation~\eqref{CAE} as well.
The corresponding surfaces~\eqref{invol} exhaust all constant astigmatism surfaces sharing one of the evolutes with the
seed surface~$\mathbf r$.
\end{proposition}

\begin{proof}
The statements concerning $x'_A$, $y'_A$ and $z'_A$ follow from formulas~\eqref{XA} in the preceding section.
Actually, the integration constants~$b$, $c_1$, $c_2$ can be combined into a~square matrix~$A$ such that $\det A = \pm
1$; namely,
\begin{gather*}
a_{11} = \frac{c_1}{\sqrt{|b|}},
\qquad
a_{12} = \frac{1}{\sqrt{|b|}},
\qquad
a_{21} = \frac{b - c_1 c_2}{\sqrt{|b|}},
\qquad
a_{22} = -\frac{c_2}{\sqrt{|b|}}.
\end{gather*}
Formulas for $x^*_A$, $y^*_A$, $z^*_A$ follow from these with the help of the involution ${\mathcal J}$, which interchanges
the evolutes.
\end{proof}

Observe that purely imaginary values $a_{ij}$ also produce a~real result.
Then the `$\pm$' sign in front of $y_A'$ and $x_A^*$ can be circumvented by combination with the reversals ${\mathcal
R}^x$, ${\mathcal R}^y$, because of the easy identities
\begin{gather*}
{\mathcal X}_{{\rm i} A} = {\mathcal R}^y \circ {\mathcal X}_A,
\qquad
{\mathcal Y}_{{\rm i} A} = {\mathcal R}^x \circ {\mathcal Y}_A.
\end{gather*}
Some useful identif\/ications are:
\begin{gather*}
\begin{array}{@{}rccc} \text{symmetry} & {\mathcal T}^x_a & {\mathcal S}_c & {\mathcal X}
\\
\hline
\text{matrix~$A$} & \left(
\begin{matrix}
{\rm i} & 0
\\
a{\rm i} & {\rm i}
\end{matrix}
\right)  &  \left(
\begin{matrix}
\sqrt c & 0
\\
0 & 1/\sqrt c
\end{matrix}
\right)  &  \left(
\begin{matrix}
0 & 1
\\
1 & 0
\end{matrix}
\right)\tsep{10pt}
\end{array}
\end{gather*}
Recall that the translation ${\mathcal T}^y_a$ is due to the non-uniqueness of~$\eta$, see Remark~\ref{rem:nonuniq}.
Otherwise said, ${\mathcal T}^y_a$ corresponds to the unit matrix.
The proofs of the following two propositions are straightforward, hence omitted.

\begin{proposition}
In the case when $a_{12} = 0$ the transformations reduce to local symmetries
\begin{gather*}
{\mathcal X}_A =
\begin{cases}
{\mathcal T}^x_{a_{21}/a_{11}} \circ {\mathcal S}_{-a_{11}^2} & \text{if} \ \det A = -1,
\\
{\mathcal T}^x_{a_{21}/a_{11}} \circ {\mathcal R}^y \circ {\mathcal S}_{-a_{11}^2} & \text{if} \ \det A = +1,
\end{cases}
\\
{\mathcal Y}_A =
\begin{cases}
{\mathcal T}^y_{a_{21}/a_{11}} \circ {\mathcal S}_{-1/a_{11}^2} & \text{if} \ \det A = -1,
\\
{\mathcal T}^y_{a_{21}/a_{11}} \circ {\mathcal R}^x \circ {\mathcal S}_{-1/a_{11}^2} & \text{if} \ \det A = +1.
\end{cases}
\end{gather*}
\end{proposition}

\begin{proposition}
We have
\begin{gather*}
{\mathcal X}_B \circ {\mathcal X}_A = {\mathcal X}_{B A},
\qquad
{\mathcal Y}_B \circ {\mathcal Y}_A = {\mathcal Y}_{B A}
\end{gather*}
for any two $2 \times 2$ matrices~$A$,~$B$ such that $|{\det A}| = |{\det B}| = 1$.
\end{proposition}

It follows that transformations~${\mathcal X}_A$ form a~three-parameter group, and similarly for the
transformations~${\mathcal Y}_A$.

\section{The reciprocal transformation as a~nonlocal symmetry}
\label{sect:nloc}

Besides the geometrical construction presented in Section~\ref{sect:Constr}, there is also a~more systematic way to
derive the transformations ${\mathcal X}_A$ and ${\mathcal Y}_B$.
Consider the system formed by the constant astigmatism equation~\eqref{CAE} and the f\/irst four
equations~\eqref{potentials}, i.e.,
\begin{gather}
z_{yy} + \left(\frac{1}{z}\right)_{xx} + 2 = 0,
\nonumber
\\
\chi_x = z_y + y,
\qquad
\chi_y = \frac{z_x}{z^2} - x,
\qquad
\eta_x = x z_y,
\qquad
\eta_y = x \frac{z_x}{z^2} + \frac1z - x^2.
\label{chietacov}
\end{gather}
According to~\cite{B-V-V}, system~\eqref{chietacov} constitutes a~{\it cove\-ring} of the constant astigmatism equation.
The Lie algebra~$\mathfrak X$ of Lie symmetries of system~\eqref{chietacov} is routinely computable.
Omitting details, we present the basis
\begin{gather*}
\frac\partial{\partial \chi},
\qquad
\frac\partial{\partial \eta},
\qquad
-z_x \frac\partial{\partial z} - (z_y + 2 y) \frac\partial{\partial \chi} - (x z_y - \chi + x y) \frac\partial{\partial
\eta},
\nonumber
\\
-z_y \frac\partial{\partial z} - \left(\frac{z_x\vphantom{1}}{z^2} - 2 x\right)\frac\partial{\partial \chi} - \left(\frac{x}{z^2}
z_x + \frac{1}{z} - x^2\right) \frac\partial{\partial \eta},
\nonumber
\\
(x z_x - y z_y + 2 z) \frac\partial{\partial z} + \left(x z_y - \frac{y\vphantom{1}}{z^2} z_x + 2 y x\right) \frac\partial{\partial
\chi} + \left(x^2 z_y - \frac{x y\vphantom{1}}{z^2} z_x - \eta + x^2 y - \frac{y}{z}\right) \frac\partial{\partial \eta},
\nonumber
\\
\left((\chi - y x) z_y + \left(x^2 - \frac{1}{z}\right) z_x + 4 x z\right) \frac\partial{\partial z}
\nonumber
\\
\qquad
{} + \left(\left(x^2 - \frac{1}{z}\right) z_y + \frac{\chi - y x}{z^2} z_x - 2 x \chi + 2 \eta + 2 y x^2\right)
\frac\partial{\partial \chi}
\nonumber
\\
\qquad
{} + \left(\left(x^2 - \frac{1}{z}\right) x z_y + \frac{\chi - y x}{z^2} x z_x - \left(x^2 -
\frac{1}{z}\right)\frac{\chi - y x}{z}\right) \frac\partial{\partial \eta}
\end{gather*}
of a~six-dimensional Lie algebra.
Adding a~suitable linear combination
of total derivatives, we transform these generators into six vector f\/ields
\begin{gather*}
{\mathfrak t}^x_{\mathfrak X} = \frac\partial{\partial x} - y \frac\partial{\partial\chi} + (\chi - x y)
\frac\partial{\partial\eta},
\qquad
{\mathfrak t}^y_{\mathfrak X} = \frac\partial{\partial y} + x \frac\partial{\partial\chi},
\nonumber
\\
{\mathfrak t}^\chi_{\mathfrak X} = \frac\partial{\partial\chi},
\qquad
{\mathfrak t}^\eta_{\mathfrak X} = \frac\partial{\partial\eta},
\qquad
{\mathfrak s}_{\mathfrak X} = -x \frac\partial{\partial x} + y \frac\partial{\partial y} + 2 z \frac\partial{\partial z}
- \eta \frac\partial{\partial\eta},
\nonumber
\\
{\mathfrak x}_{\mathfrak X} = \left(\frac1z - x^2\right)\frac\partial{\partial x} + (x y - \chi) \frac\partial{\partial
y} + 4 x z \frac\partial{\partial z} + \left(\frac{y\vphantom{1}}{z} - x \chi + 2 \eta\right) \frac\partial{\partial\chi}
\end{gather*}
acting in the f\/ive-dimensional space coordinatised by~$x$,~$y$,~$z$,~$\chi$,~$\eta$.
Their non-vanishing commutators are
\begin{alignat*}
{5} & [{\mathfrak t}^x_{\mathfrak X},{\mathfrak t}^y_{\mathfrak X}] = 2{\mathfrak t}^\chi_{\mathfrak X},
\qquad
&& [{\mathfrak t}^\chi_{\mathfrak X},{\mathfrak t}^x_{\mathfrak X}] = {\mathfrak t}^\eta_{\mathfrak X}, && && &
\\
& [{\mathfrak s}_{\mathfrak X},{\mathfrak t}^\eta_{\mathfrak X}] = {\mathfrak t}^\eta_{\mathfrak X},
\qquad
&& [{\mathfrak s}_{\mathfrak X},{\mathfrak t}^x_{\mathfrak X}] = {\mathfrak t}^x_{\mathfrak X},
\qquad
&& [{\mathfrak s}_{\mathfrak X},{\mathfrak t}^y_{\mathfrak X}] = -{\mathfrak t}^y_{\mathfrak X},
\qquad
&& [{\mathfrak s}_{\mathfrak X},{\mathfrak x}_{\mathfrak X}] = {\mathfrak x}_{\mathfrak X},&
\\
& [{\mathfrak x}_{\mathfrak X},{\mathfrak t}^\chi_{\mathfrak X}] = {\mathfrak t}^y_{\mathfrak X},
\qquad
&& [{\mathfrak x}_{\mathfrak X},{\mathfrak t}^\eta_{\mathfrak X}] = -2 {\mathfrak t}^\chi_{\mathfrak X},
\qquad
&& [{\mathfrak x}_{\mathfrak X},{\mathfrak t}^x_{\mathfrak X}] = -2 {\mathfrak s}_{\mathfrak X}.
&& &
\end{alignat*}
Alternatively speaking, ${\mathfrak X}$ is the Lie algebra of nonlocal symmetries~\cite{B-V-V} corresponding to
co\-ve\-ring~\eqref{chietacov} of the constant astigmatism equation.

The f\/lows (one-parametric groups) induced by the generators of ${\mathfrak X}$ are easy to compute.
The f\/low of ${\mathfrak t}^\chi_{\mathfrak X}$ is simply ${\mathcal T}^\chi_{\mathfrak X}: \chi \mapsto \chi + t$,
where~$t$ is the parameter.
Similarly, the f\/low of ${\mathfrak t}^\eta_{\mathfrak X}$ is simply ${\mathcal T}^\eta_{\mathfrak X}: \eta \mapsto \eta
+ t$.
These two f\/lows ref\/lect the freedom to choose the integration constants in system~\eqref{chietacov}.
Neither of them alters the solution $z(x,y)$.

Formulas for the remaining four f\/lows occupy the columns of the table
\begin{gather*}
\begin{array}
{@{}l|cccc} & {\mathcal T}^y_{\mathfrak X} & {\mathcal T}^x_{\mathfrak X} & {\mathcal S}_{\mathfrak X} & {\mathcal
X}_{\mathfrak X}\bsep{2pt}
\\
\hline
y & y + t & y & y {\rm e}^t & y + (x y - \chi) t - \eta t^2\tsep{2pt}
\\
x & x & x + t & \dfrac x{{\rm e}^t} & \dfrac{(x t + 1) x z + t}{(x t + 1)^2 z + t^2}
\\
z & z & z & z {\rm e}^{2t} & \big((x t + 1)^2 z + t^2\big)^2/z\tsep{2pt}\bsep{2pt}
\\
\chi & \chi + x t & \chi - y t & \chi & \dfrac{(x t + 1) (\eta x t^2 + 2 \eta t + \chi) z + (\eta t^2 + y) t}{(x t +
1)^2 z + t^2}
\\
\eta & \eta & \eta + (\chi - x y) t - y t^2 & \dfrac\eta{{\rm e}^t} & \eta
\end{array}
\end{gather*}
Of course, the f\/irst three columns are the~$x$-translation ${\mathcal T}^x$, the~$y$-translation~${\mathcal T}^y$, and
the sca\-ling~${\mathcal S}$, extended to the nonlocal variables~$\chi$,~$\eta$.
Finally, the rightmost column harbours an extension ${\mathcal X}_{\mathfrak X}$ of the transformation~${\mathcal X}_A$,
where
\begin{gather*}
A = \left(
\begin{matrix}
1 & t
\\
0 & 1
\end{matrix}
\right).
\end{gather*}

Another six-dimensional algebra $\mathfrak Y$ results from an analogous computation using the
poten\-tials~$\chi$,~$\xi$.
Alternatively, $\mathfrak Y$ is conjugated to ${\mathfrak X}$ by means of the involution ${\mathcal J}$.
Therefore, ${\mathfrak X} \cong {\mathfrak Y}$.
We omit the explicit description of its generators.
The construction of an inf\/inite-dimensional covering to harbour extensions of both ${\mathfrak X}$ and ${\mathfrak Y}$
is postponed to a~forthcoming paper.

\section{Relation to the sine-Gordon equation}
\label{sect:sG}

As already mentioned in Section~\ref{sect:Constr} above, every constant astigmatism surface $\mathbf r$ yields two
pseudospherical surfaces (evolutes) $\hat{\mathbf r}_1$ and $\hat{\mathbf r}_2$ (in Section~\ref{sect:Constr}, the
subscript was dropped since we considered only one of the evolutes).
The two pseudospherical surfaces $\hat{\mathbf r}_1$, $\hat{\mathbf r}_2$ are said to be {\it complementary}
(Bianchi~\cite[p.~285]{Bia-a}).
As is well known, the evolutes $\hat{\mathbf r}_i$ admit common coordinates~$\alpha$,~$\beta$ that are both asymptotic
and Chebyshev, i.e.,
\begin{gather*}
\hat{\mathbf{I}}_i = {\mathrm d}\alpha^2 + 2 \cos\phi_i \, {\mathrm d}\alpha\, {\mathrm d}\beta + {\mathrm d}\beta^2,
\qquad
\hat{\mathbf{II}}_i = 2 \sin\phi_i \, {\mathrm d}\alpha \, {\mathrm d}\beta.
\end{gather*}
Correspondingly, every solution $z(x,y)$ of the constant astigmatism equation yields two complementary solutions
$\phi_i(\alpha,\beta)$ of the sine-Gordon equation.
According to~\cite[equation~(29)]{B-M-I},~$\alpha$~and~$\beta$ are given by equations~\eqref{alphabeta}, while
formulas~\cite[equation~(30)]{B-M-I} can be simplif\/ied to
\begin{gather}
\cos\phi_1 = \frac{z_x^2 - z^2 z_y^2 + 4 z^3} {\sqrt{(z_x + z z_y)^2 + 4 z^3} \sqrt{(z_x - z z_y)^2 + 4 z^3}},
\nonumber
\\
\sin\phi_1 = \frac{4 z^{5/2} z_y} {\sqrt{(z_x + z z_y)^2 + 4 z^3} \sqrt{(z_x - z z_y)^2 + 4 z^3}},
\nonumber
\\
\cos\phi_2 = \frac{z_x^2 - z^2 z_y^2 - 4 z^3} {\sqrt{(z_x + z z_y)^2 + 4 z^3} \sqrt{(z_x - z z_y)^2 + 4 z^3}},
\nonumber
\\
\sin\phi_2 = \frac{4 z^{3/2} z_x} {\sqrt{(z_x + z z_y)^2 + 4 z^3} \sqrt{(z_x - z z_y)^2 + 4 z^3}}.
\label{sincosphi}
\end{gather}
The equalities $\phi_{i,\alpha\beta} = \sin\phi_i$ are straightforward to check.

Let us remark that formulas~\eqref{sincosphi} can be replaced with simple
\begin{gather*}
\phi_1 = \arctan\frac{4z^{3/2} z_x}{z_x^2-4 z^3-z^2 z_y^2}
\qquad
\text{and}
\qquad
\phi_2 = \arctan\frac{4 z^{5/2} z_y}{z_x^2 +4 z^3 - z^2 z_y^2},
\end{gather*}
but care must be taken as to which of the two values $\phi_i$, $\phi_i + \pi$ is to be chosen.

We obtain mappings $\mathcal F_i(x,y,z) = (\alpha,\beta,\phi_i)$ from the constant astigmatism equation to the
sine-Gordon equation.

\begin{proposition}
\label{prop:RTdiagrams}
We have $\mathcal F_1 \circ {\mathcal X}_A = \mathcal F_1$ and $\mathcal F_2 \circ {\mathcal Y}_B = \mathcal F_2$, i.e.,
the diagrams
\begin{gather*}
\xymatrix{
{(x,y,z)} \ar[rr]^{{\mathcal X}_A} \ar[rd]_{{\mathcal F}_1}
 & & {(x'_A,y'_A,z'_A)} \ar[ld]^{{\mathcal F}_1} \\
& {(\alpha,\beta,\phi_1)}
}
\qquad
\xymatrix{
{(x,y,z)} \ar[rr]^{{\mathcal Y}_B} \ar[rd]_{{\mathcal F}_2}
 & & {(x^*_B,y^*_B,z^*_B)} \ar[ld]^{{\mathcal F}_2} \\
& {(\alpha,\beta,\phi_2)}
}
\end{gather*}
commute.
\end{proposition}

\begin{proof}
By Proposition~\ref{prop:A}, the solutions $(x'_A,y'_A,z'_A) = {\mathcal X}_A(x,y,z)$ and $(x^*_B,y^*_B,z^*_B) =
{\mathcal Y}_B(x,y,z)$ exhaust all constant astigmatism solutions sharing one of the evolutes with the seed $(x,y,z)$.
The statements follow by introducing the common asymptotic-Chebyshev parameterization of the evolutes.
\end{proof}

Let us discuss the reciprocal transformation in terms of the sine-Gordon solutions.
It is closely related to the B\"acklund relation $\mathcal{B}^{(\lambda)}$, given by the system
\begin{gather}
\left(\frac{\psi - \phi}{2}\right)_{\beta} = \lambda \sin \frac{\psi + \phi}{2},
\qquad
\left(\frac{\psi + \phi}{2}\right)_{\alpha} = \frac{1}{\lambda} \sin \frac{\psi - \phi}{2}.
\label{BT}
\end{gather}
The customary notation is $\psi = \mathcal{B}^{(\lambda)}\phi$, even though the image~$\psi$ of a~given solution~$\phi$
depends on one integration constant.
Formulas~\eqref{BT} are invariant under the switch $\psi \leftrightarrow \phi$, $\lambda \leftrightarrow -\lambda$.
Therefore, relations $\mathcal{B}^{(\lambda)}$ and $\mathcal{B}^{(-\lambda)}$ are inverse one to another, meaning that
if $\psi = \mathcal{B}^{(\lambda)}\phi$ for a~particular choice of the integration constant, then $\phi =
\mathcal{B}^{(-\lambda)}\psi$ for a~particular choice of the integration constant.
It should be stressed that neither $\mathcal{B}^{(\lambda)} \circ \mathcal{B}^{(-\lambda)}$ nor
$\mathcal{B}^{(-\lambda)} \circ \mathcal{B}^{(\lambda)}$ is the identity (see Example~\ref{ex:BT}).

\begin{remark}
\label{rem:blur}
Actually, the dif\/ference between $\mathcal{B}^{(\lambda)}$ and $\mathcal{B}^{(-\lambda)}$ is somewhat blurred.
Note that solutions of the sine-Gordon equation are determined up to adding an integer multiple of $2\pi$.
However, according to formulas~\eqref{BT}, if $\psi = \mathcal{B}^{(\lambda)}\phi$, then $\psi + 2 \pi =
\mathcal{B}^{(-\lambda)}\phi$ and also $\psi = \mathcal{B}^{(-\lambda)}(\phi + 2\pi)$.
\end{remark}

\begin{example}
\label{ex:BT}
Under $\mathcal B^{(\pm1)}$, the zero solution of the sine-Gordon equation is transformed to the 1-soliton solution $4
\arctan {\rm e}^{\pm(\alpha + \beta + c_1)}$.
The latter is transformed to
\begin{gather*}
4 \arctan \frac{\cosh(\alpha + \beta + c_1)}{\pm \alpha \mp \beta + c_2}
\end{gather*}
under the inverse relation $\mathcal B^{(\mp1)}$.
This shows that $\mathcal B^{(\mp1)} \circ \mathcal B^{(\pm1)}$ is, by far, not the identity map.
The result is determined up to translations in~$\alpha$ and~$\beta$.
However, by letting $c_2 \to \infty$ we recover the zero seed.
\end{example}

When $\lambda = \pm1$, then~$\psi$ and~$\phi$ correspond to complementary surfaces, see Bianchi~\cite[\S~375]{Bia-I}.
Otherwise said, $\mathcal{B}^{(\pm1)}$ relates the two distinct evolutes of one and the same constant astigmatism
surface.
More precisely, we have the following statement.

\begin{proposition}
\label{prop:BTdiagrams}
The diagrams
\begin{gather*}
\xymatrix{
 & {(x,y,z)} \ar[ld]^{{\mathcal F}_1} \ar[rd]_{{\mathcal F}_2}
 &
 & & {(x,y,z)} \ar[ld]^{{\mathcal F}_2} \ar[rd]_{{\mathcal F}_1}
\\
 {(\alpha,\beta,\phi_1)} \ar[rr]_{{\mathcal B}^{(1)}}
 & & {(\alpha,\beta,\phi_2)}
 &  {(\alpha,\beta,\phi_2)} \ar[rr]_{{\mathcal B}^{(-1)}}
 & & {(\alpha,\beta,\phi_1)}
}
\end{gather*}
commute, up to adding an integer multiple of $2 \pi$ to $\phi_1$ or $\phi_2$, cf.\ Remark~{\rm \ref{rem:blur}}.
\end{proposition}

\begin{proof}
Let us check formulas~\eqref{BT}, where we substitute $\phi_1$ for~$\phi$ and $\phi_2$ for~$\psi$.
Using formulas~\eqref{sincosphi}, we compute
\begin{gather*}
\frac{\phi_{1,\beta} - \phi_{2,\beta}}{2} = -\frac{2 z^{3/2}}{\sqrt{(z_x - z z_y)^2 + 4 z^3}},
\qquad
\frac{\phi_{1,\alpha} + \phi_{2,\alpha}}{2} = -\frac{2 z^{3/2}}{\sqrt{(z_x + z z_y)^2 + 4 z^3}},
\end{gather*}
which gives the left-hand side.
Next we compute
\begin{gather*}
\sin^2\frac{\phi_2 \pm \phi_1} 2 = \frac{1 - \cos(\phi_1 \pm \phi_2)} 2 = \frac{1 \pm \sin\phi_1 \sin \phi_2 -
\cos\phi_1 \cos \phi_2} 2 = \frac{4 z^3}{(z_x \mp z z_y)^2 + 4 z^3},
\end{gather*}
from where we can reconstruct the right-hand side, up to a~sign.
A~mismatch of the signs can be rectif\/ied by adding an integer multiple of~$2 \pi$ to $\phi_2$ or $\phi_1$.
\end{proof}

\begin{figure}[ht]\centering
$ \xymatrix{\bullet \ar[rr]_{{\mathcal X}_{A}} \ar[rd]_{\mathcal F_1} & & \bullet \ar[ld]^{\mathcal F_1}
\ar[rr]_{{\mathcal Y}_{B}} \ar[rd]_{\mathcal F_2} & & \bullet \ar[ld]^{\mathcal F_2} \ar[rr]_{{\mathcal X}_{A}}
\ar[rd]_{\mathcal F_1} & & \bullet \ar[ld]^{\mathcal F_1} \ar[rr]_{{\mathcal Y}_{B}} \ar[rd]_{\mathcal F_2} & & \bullet
\ar[ld]^{\mathcal F_2} \ar[rd]_{\mathcal F_1}
\\
& \bullet \ar[rr]_{\mathcal B^{(1)}} & & \bullet \ar[rr]_{\mathcal B^{(-1)}} & & \bullet \ar[rr]_{\mathcal B^{(1)}} & &
\bullet \ar[rr]_{\mathcal B^{(-1)}} & & \bullet} $
\caption{Connection between reciprocal transformations for the
constant astigmatism equation and the B\"acklund transformation for the sine-Gordon equation.}
\label{BT1}
\end{figure}
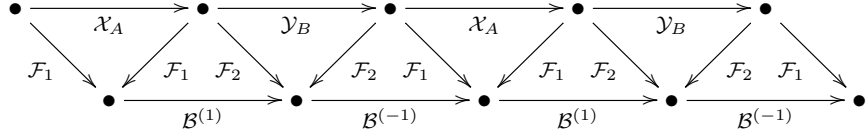

Combining Propositions~\ref{prop:BTdiagrams} and~\ref{prop:RTdiagrams}, we obtain a~commutative diagram, see
Fig.~\ref{BT1}, inf\/initely extensible to the left and to the right.
It follows that, on the level of sine-Gordon solutions, alternate repeating of transformations ${\mathcal X}_A$ and
${\mathcal Y}_B$ corresponds to alternate repeating the B\"acklund transformation with B\"acklund parameter $\lambda =
\pm1$.

As is well known (see~\cite[Corollary~3.4 and Section~4.2]{MM} for a~geometric proof), the B\"acklund transformation of
the sine-Gordon equation, applied repeatedly, produces solutions depending on an ever increasing number of integration
constants.
Consequently, transformations~${\mathcal X}$ and~${\mathcal Y}$, applied repeatedly, produce solutions of the constant
astigmatism equation, depending on an ever increasing number of arbitrary parameters.

\section{Transformation of constant astigmatism surfaces
\\
and orthogonal equiareal patterns}
\label{sect:OEP}

Let $x$, $y$ be the coordinates the constant astigmatism equation is referred to and let $z(x,y)$ be its solution.
Then the third fundamental form (the metric on the Gaussian sphere) of the corresponding constant astigmatism surface is
$\mathbf{III} = z\,{\mathrm d} x^2 + (1/z)\,{\mathrm d} y^2$, i.e., we obtain an orthogonal equiareal pattern (see the
Introduction) on the Gaussian sphere.

The Gaussian image, $\tilde{\mathbf n}$, of the transformed surface is given by formula~\eqref{normal}.
In combination with the last line of equations~\eqref{GW} one obtains
\begin{gather}
\label{tilde n}
\tilde{\mathbf{n}} = \frac{x^2 z - 1}{x^2 z + 1}\mathbf{n} + \frac{2x}{x^2 z + 1} \mathbf{n}_x.
\end{gather}
It is easily checked that the f\/irst fundamental form of $\tilde{\mathbf{n}}$ in terms of coordinates $x'$, $y'$ def\/ined
by~\eqref{RTX} is
\begin{gather*}
\mathbf{I}_{\tilde{\mathbf{n}}} = z'\, {\mathrm d} x'^2 + \frac{1}{z'}\, {\mathrm d} y'^2
\end{gather*}
and therefore generates a~new orthogonal equiareal pattern on the transformed surface's Gaussian sphere.

What is the relationship between the initial and the transformed pattern? Let~$\psi$ denote the angle between
$\mathbf{n}$ and $\tilde{\mathbf n}$ (not to be confused with the~$\psi$ of the preceding section).
Then, according to~\eqref{normal},
\begin{gather*}
\cos\psi = \frac{x^2 z - 1}{x^2 z + 1}
\end{gather*}
and, therefore,
\begin{gather}
\label{cothalfangle}
\cot \frac{\psi}{2} = x\sqrt{z},
\end{gather}
where $x\sqrt z$ is the invariant of the reciprocal transformation introduced in Remark~\ref{rem:circinversion}.
The angle~$\psi$ can be determined by formula~\eqref{cothalfangle} up to an integer multiple of~$2\pi$; then $\cos\psi$
is as above and
\begin{gather*}
\sin\psi = \frac{2 x \sqrt z}{x^2 z + 1}.
\end{gather*}
Then formula~\eqref{tilde n} can be rewritten as
\begin{gather*}
\tilde{\mathbf{n}} = \cos\psi \, \mathbf{n} + \sin\psi \, \frac{\mathbf{n}_x}{\sqrt z},
\end{gather*}
where $\mathbf{n}_x/\sqrt z$ is the unit vector codirectional with $\mathbf{n}_x$.

The vectors tangent to the lines $y' = {\rm const}$ and $x' = {\rm const}$ at the point $\tilde{\mathbf{n}}(x,y)$ are
\begin{gather*}
\tilde{\mathbf n}_{x'} = \frac{x^2z - 1}{z} \mathbf{n}_x - 2 x \mathbf{n},
\qquad
\tilde{\mathbf n}_{y'} = -\frac{z}{x^2z + 1}\mathbf{n}_y.
\end{gather*}
Consequently, $\mathbf n$, $\mathbf n_x$, $\tilde{\mathbf n}$ and $\tilde{\mathbf n}_{x'}$ lie in one and the same plane,
while $\mathbf{n}_y$ and $\tilde{\mathbf n}_{y'}$ are ortho\-go\-nal to it.
The angle between $\tilde{\mathbf{n}}_{x'}$ and $\mathbf{n}_x$ is~$\psi$.

\begin{figure}[ht] \centering
\includegraphics[width=5.0cm]{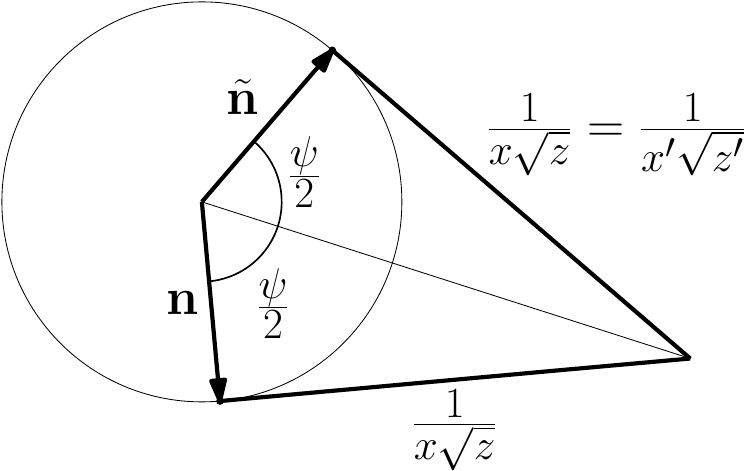}
\caption{The transformation of an orthogonal equiareal pattern.
Intersection of the Gaussian sphere with the plane containing $\mathbf n$, $\mathbf n_x$, $\tilde{\mathbf n}$,
$\tilde{\mathbf n}_{x'}$.}
\label{OEP}
\end{figure}

The transformed orthogonal equiareal pattern can be constructed in the following way: Rotate the vector $\mathbf{n}$~by
angle~$\psi$ in the plane spanned by $\mathbf{n}$ and $\mathbf{n}_x$.
One of the new tangent vectors, $\tilde{\mathbf{n}}_{x'}$, lies in the above-mentioned plane while the second one,
$\tilde{\mathbf{n}}_{y'}$, is orthogonal to it.
Fig.~\ref{OEP} provides a~schematic picture of the construction.

Similarly, formula~\eqref{invol}, which describes the reciprocal transformation in terms of constant astigmatism
surfaces, can be rewritten simply as
\begin{gather*}
\tilde{\mathbf{r}} = \mathbf{r} + (\sigma - \sigma' \cos \psi) \mathbf{n} - \sigma' \sin \psi\,  \mathbf{e},
\end{gather*}
where $\sigma'$ is given by formula~\eqref{sigmaprime} and $\mathbf e = -\mathbf{r}_x/u$ is a~unit vector codirectional
or contradirectional (depending on the value of~$z$) with $\mathbf{r}_x$.
Clearly, we can rewrite $\mathbf{r}$ as the dif\/ference of $\mathbf{r} + \sigma \mathbf{n}$, which is the evolute, and
$(\cos \psi \, \mathbf{n} + \sin \psi\,  \mathbf{e}) \sigma'$, which is the evolute of the transformed surface.

\section{Examples}
\label{section8}

\begin{example}
\rm
\label{ex:vL}
Let us apply the transformations ${\mathcal X}$ and ${\mathcal Y}$ to the von Lilienthal solution
\begin{gather*}
z = b^2 - y^2,
\end{gather*}
where~$b$ is a~constant.
The name comes from the fact that~$z$ corresponds to surfaces studied by von Lilienthal~\cite{Lil}, see~\cite{B-M-I}.

Using~\eqref{RTX}, we obtain ${\mathcal X}(x,y,z) = (x', y', z')$, where
\begin{gather}
x' = \frac{x \big(b^2-y^2\big)}{x^2\big(b^2-y^2\big) + 1},
\nonumber
\\
y' = \eta = \int x z_y \, {\mathrm d} x + \left(x \frac{z_x}{z^2} + \frac1z - x^2\right) {\mathrm d} y = \frac1 {b}
\mathrm{arctanh}\left(\frac{y}{b}\right) - x^2y + c_1,
\nonumber
\\
z' = \frac{(x^2 \big(b^2-y^2\big) + 1)^2}{b^2-y^2},
\label{XvL}
\end{gather}
$c_1$ being the integration constant.
Here~$\eta$ has been expressed as a~path-independent line integral according to formula~\eqref{potentials}.
Apparently, $z'(x',y')$ is a~substantially new solution of the equation~\eqref{CAE}.

Similarly, using~\eqref{RTY}, we obtain ${\mathcal Y}(x,y,z) = (x^*, y^*, z^*)$, where
\begin{gather*}
x^* = b^2x + c_2,
\qquad
y^* = \frac{y}{b^2},
\qquad
z^* = \frac{b^2-y^2}{b^4}.
\end{gather*}
However, $z^* = -{y^*}^2 + 1/b^2$ and, thus, we obtained just another von Lilienthal solution.
\end{example}

\begin{remark}
Examples in this section demonstrate that reciprocal transformations inevitably produce solutions in parametric form.
While inconvenient, this is not a~serious obstacle.
Both iteration of the procedure and construction of the constant astigmatism surface or the orthogonal equiareal pattern
are possible.
However, it is not straightforward to see whether two solutions coincide up to a~reparameterization.
\end{remark}

\begin{example}
The general von Lilienthal solution $z = -y^2 + k y + l$ is related to $z = b^2 - y^2$ by a~$y$-translation.
To obtain its ${\mathcal X}$-transformation one can employ the identity ${\mathcal X} \circ {\mathcal T}^y_a = {\mathcal
T}^y_a \circ {\mathcal X}$, while its ${\mathcal Y}$-transformation is one of the von Lilienthal solutions again.
\end{example}

\begin{example}
Continuing Example~\ref{ex:vL}, we apply transformation ${\mathcal Y}$ to the solution $z'(x',y')$.
According to~\eqref{RTY},
\begin{gather*}
x^{\prime*} = \xi',
\qquad
y^{\prime*} = \frac{y'}{y^{\prime2} + z'},
\qquad
z^{\prime*} = \frac{z'}{\big(y^{\prime2} + z'\big)^2}.
\end{gather*}
It is a~matter of algebraic manipulations to compute
\begin{gather*}
y^{\prime*} = \frac{\dfrac1b \operatorname{arctanh}\left(\dfrac{y}{b}\right) - x^2 y} {\left(\dfrac1b
\operatorname{arctanh}\left(\dfrac{y}{b}\right) - x^2 y\right)^2 + \dfrac{\big(\big(b^2 - y^2\big) x^2 + 1\big)^2}{b^2 -
y^2}},
\\
z^{\prime*} = \frac{\big(\big(b^2 - y^2\big) x^2 + 1\big)^2} {\big(b^2 - y^2\big) \left(\left(\dfrac1b
\operatorname{arctanh}\left(\dfrac{y}{b}\right) - x^2 y\right)^2 + \dfrac{\big(\big(b^2 - y^2\big) x^2 + 1\big)^2}{b^2 -
y^2}\right)^{2}}.
\end{gather*}
Omitting details, we compute $x^{\prime*}$ as the path-independent line integral
\begin{gather*}
x^{\prime*} = \xi' = \int -\big(y' z'_{y'} - z' + y^{\prime2}\big)\,  {\mathrm d} x' - y' \frac{z'_{x'}}{z^{\prime2}}
\, {\mathrm d} y'
\\
\hphantom{x^{\prime*}}{} = - \frac{\big(b^2 - y^2\big) x}{\big(b^2 - y^2\big) x^2 + 1} \left(\frac1b
\operatorname{arctanh}\left(\frac{y}{b}\right) - x^2 y\right)^2
\\
\hphantom{x^{\prime*} =}{} -2 x y \left(\frac1b \operatorname{arctanh}\left(\frac{y}{b}\right) - x^2 y\right) - x
\frac{\big(b^2 + 3y^2\big) x^2 - 3}{3}.
\end{gather*}
Thus, we have obtained one more solution in a~parametric form.
\end{example}

\begin{example}
Continuing Example~\ref{ex:vL}, we provide a~picture of the surface of constant astigmatism generated from the von
Lilienthal seed by transformation ${\mathcal X}$.
The von Lilienthal surfaces are obtained by revolving the involutes of tractrix around the asymptote of the latter,
see~\cite{B-M-I} for pictures.
We can write
\begin{gather*}
\mathbf{r}_1 = \frac1{2b} \sqrt{b^2 - y^2} \big[2 - \ln\big(b^2 - y^2\big)\big] \cos b x,
\qquad
\mathbf{r}_2 = \frac1{2b} \sqrt{b^2 - y^2} \big[2 - \ln\big(b^2 - y^2\big)\big] \sin b x,
\\
\mathbf{r}_3 = \frac12 \left(\frac yb + 1\right) \ln\big(b^2 - y^2\big) - \ln(b - y) - \frac yb,
\end{gather*}
and
\begin{gather}
\label{nvL}
\mathbf{n}_1 = \frac{\sqrt{b^2 - y^2}}{b} \cos b x,
\qquad
\mathbf{n}_2 = \frac{\sqrt{b^2 - y^2}}{b} \sin b x,
\qquad
\mathbf{n}_3 = -\frac yb.
\end{gather}
From~\eqref{invol} we obtain a~formula for $\tilde{\mathbf{r}}$.
For brevity we present it with the of\/fsetting parameter~$a$ set to zero:
\begin{gather*}
\tilde{\mathbf{r}}_1 = \gamma(b,x,y) \big\{2 b \sin(b x) - \big[x^2 \big(b^2 - y^2\big) - 1\big] \cos(b x)\big\},
\\
\tilde{\mathbf{r}}_2 = -\gamma(b,x,y) \big\{2 b \cos(b x) + \big[x^2 \big(b^2 - y^2\big) - 1\big] \sin(b x)\big\},
\\
\tilde{\mathbf{r}}_3 = \frac{x^2 \big(b^2 - y^2\big) - 1}{x^2 \big(b^2 - y^2\big) + 1} y \ln\big[x^2 \big(b^2 - y^2\big)
+ 1\big]
\\
\hphantom{\tilde{\mathbf{r}}_3 =}{} + \frac{b + y}{2 b} \frac{x^2 (b - y)^2 + 1}{x^2 \big(b^2 - y^2\big) + 1} y
\ln\big(b^2 - y^2\big) - \ln(b - y) - \frac{y}{b},
\end{gather*}
where
\begin{gather*}
\gamma(b,x,y) = \frac{\sqrt{b^2 - y^2}}{2 b} \frac{x^2 \big(b^2 - y^2\big) + 1} {2\ln\big[x^2 \big(b^2 - y^2\big) +
1\big] - \ln\big(b^2 - y^2\big) + 1}.
\end{gather*}
Obviously, $\tilde{\mathbf{r}}$ is real only if $-b < y < b$.
It is easy to check that $\tilde{\mathbf{r}}$ develops a~singularity (cuspidal edge) if either
\begin{gather*}
x^2 = \delta(a,b,y)
\qquad
\text{or}
\qquad
x^2 = \delta(a+1,b,y),
\end{gather*}
where
\begin{gather*}
\delta(a,b,y) = \frac{-1 + {\rm e}^{a} \sqrt{b^2 - y^2}}{b^2 - y^2}.
\end{gather*}
A~part of the surface $\tilde{\mathbf{r}}$ for $x > \sqrt{\delta(a+1,b,y)}$ is shown on Fig.~\ref{XvLb} under the
parameterization by~$x$,~$y$.
The cuspidal edge is clearly seen.
To parameterize the surface by lines of curvature, one would have to express~$x$,~$y$ in terms of~$x'$,~$y'$ from
formula~\eqref{XvL}.
\begin{figure}[ht] \centering
\includegraphics[width=5.2cm]{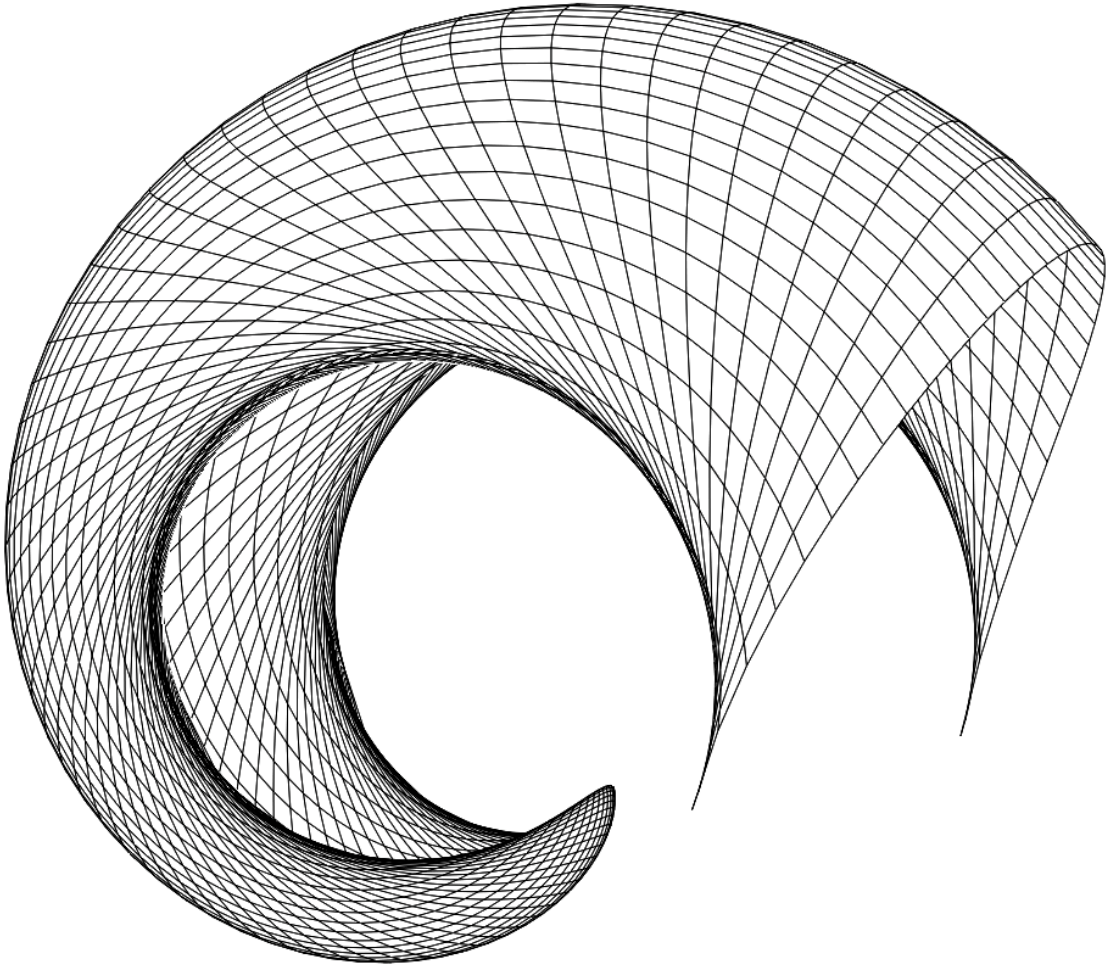}
\caption{A transformed von Lilienthal surface.}
\label{XvLb}
\end{figure}
\end{example}

\begin{example}
Continuing the previous example, we describe the transformation of the corresponding orthogonal equiareal pattern.
Under parameterization~\eqref{nvL}, the von Lilienthal solution $z = b^2 - y^2$ generates the orthogonal equiareal
pattern{\samepage
\begin{gather*}
\mathbf{I}_{\mathrm{Arch}} = \big(b^2 - y^2\big)\, {\mathrm d} x^2 + \frac{1}{b^2 - y^2}\, {\mathrm d} y^2
\end{gather*}
known as the \textit{Archimedean projection}.}

The Gaussian image of the transformed surface is
\begin{gather*}
\tilde{\mathbf{n}}_1 = \frac{\sqrt{b^2 - y^2}}{b} \frac{\big(x^2\big(b^2 - y^2\big)-1\big)\cos(bx) - 2 x b \sin(bx)}
{x^2\big(b^2 - y^2\big) + 1},
\\
\tilde{\mathbf{n}}_2 = \frac{\sqrt{b^2 - y^2}}{b} \frac{(x^2\big(b^2 - y^2\big)-1)\sin(bx) + 2xb\cos(b x)} {x^2\big(b^2
- y^2\big) + 1},
\\
\tilde{\mathbf{n}}_3 = -\frac yb \frac{x^2 \big(b^2 - y^2\big) - 1}{x^2\big(b^2 - y^2\big) + 1}.
\end{gather*}

To express the ${\mathcal X}$-transformed orthogonal equiareal pattern explicitly, one needs to invert the
transformation $(x,y) \leftrightarrow (x',y')$, where $x'$, $y'$ are given by formula~\eqref{XvL}.
The ${\mathcal X}$-transformed orthogonal equiareal pattern $\tilde{\mathbf{n}}'(x',y')$ can be seen in the right part
of Fig.~\ref{OEPvL}.
\begin{figure}[ht] \centering
\includegraphics[width=5.5cm]{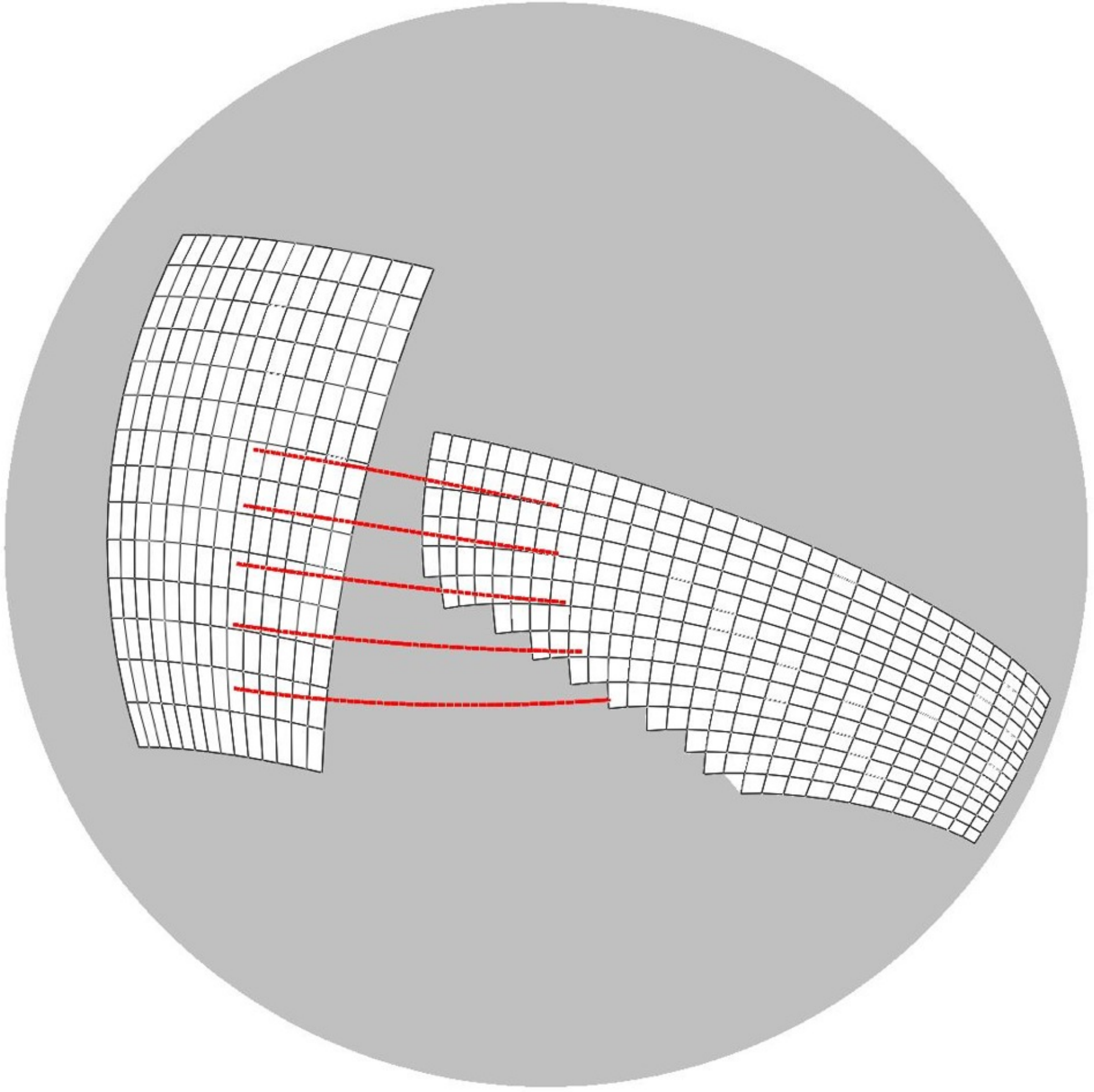}
\caption{A part of Archimedean projection (left) and a~part of
its ${\mathcal X}$-transformed pattern (right) connected by great circles' arcs (red).}
\label{OEPvL}
\end{figure}
\end{example}

\begin{example}
A~solution $z(x,y)$ is invariant under the scaling symmetry if
\begin{gather*}
x z_x - y z_y + 2 z = 0.
\end{gather*}
Consider the scaling invariant solutions
\begin{gather*}
z = \frac{F}{x^2},
\qquad
\frac{x^2 y^2}{F} + x^2 y^2 + F = a_1 x y + a_0
\end{gather*}
obtained in~\cite[proof of Proposition~1]{H-M-2}.
Recall that the corresponding surface of constant astigmatism belongs to the Lipschitz class~\cite{Lip}.
Since
\begin{gather*}
F = -\frac{1}{2} x^2 y^2 + \frac{1}{2} a_1 x y + \frac{1}{2} a_0 + \frac{1}{2} \sqrt{x^4 y^4 - 2 a_1 x^3 y^3 +
\big(a_1^2 - 2 a_0 - 4\big) x^2 y^2 + 2 a_0 a_1 x y + a_0^2}
\end{gather*}
computation of~$\eta$ and $\xi$ will lead to elliptic integrals unless the discriminant
\begin{gather*}
256 a_0^2\big(a_1^2 + 4 a_0 + 4 a_1 + 4\big)\big(a_1^2 + 4 a_0 - 4 a_1 + 4\big)
\end{gather*}
of the polynomial under the square root is zero, i.e., unless $a_0 = 0$ or $a_0 = -\big(\frac12 a_1 \pm 1\big)^2$.
To keep things simple, we restrict ourself to the easiest subcase $a_1 = a_0 = 0$.
Then
\begin{gather*}
z =\frac {y \big(\sqrt {{x}^{2}{y}^{2}-4} - xy \big)}{2x}.
\end{gather*}
The potentials are
\begin{gather*}
\chi = \sqrt{x^2y^2-4} + \arctan\frac{2}{\sqrt{x^2y^2-4}}+c_1,
\qquad
\eta = \frac{1}{2}x\big(\sqrt{x^2y^2-4}-x y \big)+c_2.
\end{gather*}
Applying the transformation ${\mathcal X}$, we obtain
\begin{gather*}
x' = \frac{1}{2} x y^2 + \frac{1}{2} y \sqrt{x^2 y^2 - 4},
\qquad
y' = -\frac{1}{2} x^2 y + \frac{1}{2} x \sqrt{x^2 y^2 - 4} + c_2,
\\
z' = -\frac{1}{2} x^4 y^2 + \frac{1}{2} \left(x^3 y - \frac{x}{y}\right) \sqrt{x^2 y^2 - 4} + \frac{3}{2} x^2.
\end{gather*}
Is is easy to check that $z'(x',y')$ satisf\/ies $x' z'_{x'} - y' z'_{y'} + 2 z' + 2 c_2 z'_{y'} = 0$, i.e., is invariant
under a~combination of the scaling symmetry and the~$y$-translation.
As such, $z'(x',y')$ is just another Lipschitz solution.

Applying the transformation ${\mathcal X}_{\mathfrak X}$, see Section~\ref{sect:nloc}, we obtain the solution
\begin{gather*}
x' = \frac{(t x + 1)\big({-}x y + \sqrt{x^2 y^2 - 4}\big) y + 2 t x} {(t x + 1)^2\big({-}x y + \sqrt{x^2 y^2 - 4}\big) y
+ 2 t^2 x},
\\
y' = -(t x + 1) y + \left(-\frac{1}{2} x^2 y + \frac{1}{2} x \sqrt{x^2 y^2 - 4} + c_2\right) t^2
\\
\hphantom{y' =}{} + \left(\sqrt{x^2 y^2 - 4} + \arctan\frac{2}{\sqrt{x^2 y^2 - 4}} + c_1\right) t,
\\
z' = \frac{(t x + 1)^4\big(\big({-}x y + \sqrt{x^2 y^2 - 4}\big) y + 2 t^2 x^2\big)^2} {x^3 y \big({-}x y + \sqrt{x^2
y^2 - 4}\big)}.
\end{gather*}
This solution is not invariant under any local symmetry.
\end{example}

\subsection*{Acknowledgements}

We are indebted to I.S.~Krasil'shchik for reading the manuscript and valuable comments.
A.H.\ was supported by Silesian University in Opava under the student grant project SGS/1/2011, M.M.\ was supported by GA\v{C}R under project P201/11/0356.

\pdfbookmark[1]{References}{ref}
\LastPageEnding


\begin{thebibliography}{99}
\footnotesize \itemsep=0pt

\bibitem{Bae}
B\"acklund A.V., Om ytor med konstant negativ kr\"okning, \textit{Lunds Univ.
  {\AA}rsskrift} \textbf{19} (1883), 1--48.

\bibitem{B-M-I}
Baran H., Marvan M., On integrability of {W}eingarten surfaces: a forgotten
  class, \href{http://dx.doi.org/10.1088/1751-8113/42/40/404007}{\textit{J.~Phys.~A: Math. Theor.}} \textbf{42} (2009), 404007,
  16~pages, \href{http://arxiv.org/abs/1002.0989}{arXiv:1002.0989}.

\bibitem{B-M-II}
Baran H., Marvan M., Classification of integrable {W}eingarten surfaces
  possessing an {$\mathfrak{sl}(2)$}-valued zero curvature representation,
  \href{http://dx.doi.org/10.1088/0951-7715/23/10/013}{\textit{Nonlinearity}} \textbf{23} (2010), 2577--2597, \href{http://arxiv.org/abs/1002.0992}{arXiv:1002.0992}.

\bibitem{Bia-a}
Bianchi L., Ricerche sulle superficie elicoidali e sulle superficie a curvatura
  costante, \textit{Ann. Scuola Norm. Sup. Pisa Cl. Sci.} \textbf{2} (1879),
  285--341.

\bibitem{Bia-I}
Bianchi L., Lezioni di Geometria Differenziale, Vol.~I, E.~Spoerri, Pisa, 1902.

\bibitem{Bia-II}
Bianchi L., Lezioni di Geometria Differenziale, Vol.~II, E.~Spoerri, Pisa,
  1903.

\bibitem{B-V-V}
Bocharov A.V., Chetverikov V.N., Duzhin S.V., Khor'kova N.G., Krasil'shchik
  I.S., Samokhin~A.V., Torkhov~Yu.N., Verbovetsky~A.M., Vinogradov~A.M.,
  Symmetries and conservation laws for differential equations of mathematical
  physics, \textit{Translations of Mathematical Monographs}, Vol.~182, Amer.
  Math. Soc., Providence, RI, 1999.

\bibitem{Fer-I}
Ferapontov E.V., Reciprocal transformations and their invariants,
  \textit{Differ. Equ.} \textbf{25} (1989), 898--905.

\bibitem{Fer-II}
Ferapontov E.V., Autotransformations with respect to the solution, and
  hydrodynamic symmetries, \textit{Differ. Equ.} \textbf{27} (1991), 885--895.

\bibitem{F-R-S}
Ferapontov E.V., Rogers C., Schief W.K., Reciprocal transformations of
  two-component hyperbolic systems and their invariants, \href{http://dx.doi.org/10.1006/jmaa.1998.6131}{\textit{J.~Math. Anal.
  Appl.}} \textbf{228} (1998), 365--376.

\bibitem{G-M}
Ganchev G., Mihova V., On the invariant theory of {W}eingarten surfaces in
  {E}uclidean space, \href{http://dx.doi.org/10.1088/1751-8113/43/40/405210}{\textit{J.~Phys.~A: Math. Theor.}} \textbf{43} (2010),
  405210, 27~pages, \href{http://arxiv.org/abs/0802.2191}{arXiv:0802.2191}.

\bibitem{Gour}
Goursat E., Le Probl\'eme de B\"acklund, Gauthier-Villars, Paris, 1925.


\bibitem{H-M}
Hlav{\'a}{\v{c}} A., Marvan M., Another integrable case in two-dimensional
  plasticity, \href{http://dx.doi.org/10.1088/1751-8113/46/4/045203}{\textit{J.~Phys.~A: Math. Theor.}} \textbf{46} (2013), 045203,
  15~pages.

\bibitem{H-M-2}
Hlav{\'a}{\v{c}} A., Marvan M., On Lipschitz solutions of the constant
  astigmatism equation, \href{http://dx.doi.org/10.1016/j.geomphys.2014.05.020}{\textit{J.~Geom. Phys.}}, to appear.

\bibitem{Ho-Mi}
Hoenselaers C.A., Miccich{\`e} S., Transcendental solutions of the
  sine-{G}ordon equation, in B\"acklund and {D}arboux Transformations. {T}he
  Geometry of Solitons ({H}alifax, {NS}, 1999), \textit{CRM Proc. Lecture
  Notes}, Vol.~29, Amer. Math. Soc., Providence, RI, 2001, 261--271.

\bibitem{K-R}
Kingston J.G., Rogers C., Reciprocal {B}\"acklund transformations of
  conservation laws, \href{http://dx.doi.org/10.1016/0375-9601(82)90081-0}{\textit{Phys. Lett.~A}} \textbf{92} (1982), 261--264.

\bibitem{Lil}
von~Lilienthal R., Bemerkung \"uber diejenigen {F}l\"achen bei denen die
  {D}ifferenz der {H}auptkr\"ummungsradien constant ist, \href{http://dx.doi.org/10.1007/BF02418056}{\textit{Acta Math.}}
  \textbf{11} (1887), 391--394.

\bibitem{Lip}
Lipschitz R., Zur {T}heorie der krummen {O}berfl\"achen, \href{http://dx.doi.org/10.1007/BF02393697}{\textit{Acta Math.}}
  \textbf{10} (1887), 131--136.

\bibitem{M-P}
Manganaro N., Pavlov M.V., The constant astigmatism equation. {N}ew exact
  solution, \href{http://dx.doi.org/10.1088/1751-8113/47/7/075203}{\textit{J.~Phys.~A: Math. Theor.}} \textbf{47} (2014), 075203,
  8~pages, \href{http://arxiv.org/abs/1311.1136}{arXiv:1311.1136}.

\bibitem{MM}
Marvan M., Some local properties of {B}\"acklund transformations, \href{http://dx.doi.org/10.1023/A:1006037726082}{\textit{Acta
  Appl. Math.}} \textbf{54} (1998), 1--25.

\bibitem{P-Z}
Pavlov M.V., Zykov S.A., Lagrangian and {H}amiltonian structures for the
  constant astigmatism equation, \href{http://dx.doi.org/10.1088/1751-8113/46/39/395203}{\textit{J.~Phys.~A: Math. Theor.}} \textbf{46}
  (2013), 395203, 6~pages, \href{http://arxiv.org/abs/1212.6239}{arXiv:1212.6239}.

\bibitem{Po-Za}
Polyanin A.D., Zaitsev V.F., Handbook of exact solutions for ordinary
  differential equations, 2nd ed., Chapman \& Hall/CRC, Boca Raton, FL, 2003.

\bibitem{P-S}
Prus R., Sym A., Rectilinear congruences and {B}\"acklund transformations:
  roots of the soliton theory, in Nonlinearity \& Geometry, Luigi Bianchi Days
  ({W}arsaw, 1995), Editors D.~W\'ojcik, J.~Cie\'sli\'nski, Polish Scientific
  Publishers, Warsaw, 1998, 25--36.

\bibitem{Ri}
Ribaucour A., Note sur les d\'evelopp\'ees des surfaces, \textit{C.~R.~Math.
  Acad. Sci. Paris} \textbf{74} (1872), 1399--1403.

\bibitem{R-S}
Rogers C., Schief W.K., B\"acklund and {D}arboux transformations. Geometry and
  modern applications in soliton theory, \href{http://dx.doi.org/10.1017/CBO9780511606359}{\textit{Cambridge Texts in Applied
  Mathematics}}, Cambridge University Press, Cambridge, 2002.

\bibitem{R-S-S}
Rogers C., Schief W.K., Szereszewski A., Loop soliton interaction in an
  integrable nonlinear telegraphy model: reciprocal and {B}\"acklund
  transformations, \href{http://dx.doi.org/10.1088/1751-8113/43/38/385210}{\textit{J.~Phys.~A: Math. Theor.}} \textbf{43} (2010),
  385210, 16~pages.

\bibitem{R-Sh}
Rogers C., Shadwick W.F., B\"acklund transformations and their applications,
  \textit{Mathematics in Science and Engineering}, Vol.~161, Academic Press,
  Inc., New York~-- London, 1982.

\bibitem{R-W}
Rogers C., Wong P., On reciprocal {B}\"acklund transformations of inverse
  scattering schemes, \href{http://dx.doi.org/10.1088/0031-8949/30/1/003}{\textit{Phys. Scripta}} \textbf{30} (1984), 10--14.

\bibitem{Sad2}
Sadowsky M.A., Equiareal pattern of stress trajectories in plane plastic
  strain, \textit{J.~Appl. Mech.} \textbf{8} (1941), A74--A76.

\bibitem{Sad1}
Sadowsky M.A., Equiareal patterns, \href{http://dx.doi.org/10.2307/2303990}{\textit{Amer. Math. Monthly}} \textbf{50}
  (1943), 35--40.



\bibitem{Wea}
Weatherburn C.E., Differential geometry of three dimensions, Cambridge
  University Press, Cambridge, 1927.

\end{thebibliography}
\end{document}